\documentclass[12pt,a4paper]{article}
\pdfoutput=1
\usepackage{a4wide}
\usepackage[centertags]{amsmath}
\usepackage{amssymb}
\usepackage{amsfonts}
\usepackage{bbm}
\usepackage{cite}
\usepackage{ulem}
\usepackage{float}
\usepackage[pdftex]{graphicx,color} 
\begin{document}
\allowdisplaybreaks

\thispagestyle{empty}
\begin{center}
{\Large  \bf  Yukawa couplings from magnetized D-brane
  models on non-factorisable tori } 
\end{center}

\vspace*{1cm}

\centerline{Stefan F{\"o}rste and Christoph Liyanage}

\vspace{1cm}

\begin{center}{\it
Bethe Center for Theoretical Physics\\
{\footnotesize and}\\
Physikalisches Institut der Universit\"at Bonn,\\
Nussallee 12, 53115 Bonn, Germany}
\end{center}

\vspace*{1cm}

\centerline{\bf Abstract}
\vskip .3cm

We compute Yukawa couplings in type IIB string theory compactified on a non 
factorisable six-torus in the presence of D9 branes and fluxes. The setting studied in
detail, is obtained by T-dualising an intersecting brane configuration of type IIA 
theory compactified on a torus generated by the SO(12) root lattice. Particular deformations of such torus are taken into account and provide moduli dependent couplings. Agreement with the type IIA result is found in a non trivial way. The classical type IIB calculation gives also information on a factor accessible only by quantum computations 
on the type IIA side.

\vskip .3cm

\newpage

\section{Introduction}

One possible extension of the Standard Model of particle physics is to assume the existence of 
extra dimensions as motivated by string theory. The appeal of such extensions lies in their
capability to explain patterns in the Standard Model which are adjusted by hand to match 
observations. One such example is the hierarchy in the size of Yukawa couplings.
In \cite{Cremades:2004wa} super Yang-Mills theory with extra dimensions was studied in this context. 
Standard Model fields correspond to zero modes of the extra dimensional Dirac equation. Different 
fields have different localisations within the extra dimensions. Yukawa couplings arise as overlap integrals of these zero modes; they are large if they are localised near to each other and small
otherwise. The authors of \cite{Cremades:2004wa} mainly focused on the extra dimensions being 
compactified on a torus which factorises into a product of two-tori. An initial $U(N)$ gauge group is broken by fluxes to $U(N_a)\times U(N_b)\times U(N_c)$ which can be further broken by Wilson lines. 
(The unbroken gauge group could e.g.\ be the Standard Model gauge group.) Computations in \cite{Cremades:2004wa} are restricted to the case that $N_a$, $N_b$, $N_c$ are mutually coprime. 
In the present paper, the discussion will be extended to particular non-factorisable tori. This will also make it necessary to abandon the restriction of $N_a$, $N_b$, $N_c$ being mutually coprime, and hence the generalisation considered
in \cite{Antoniadis:2009bg} neither applies. 

Restricting considerations to type II string model building, the above setting corresponds to type IIB theory, whereas most of type II string model building has been carried out on the type IIA side in the geometrically intuitive intersecting brane picture, see {e.g.} \cite{Berkooz:1996km,Blumenhagen:1999ev,Forste:2000hx,Forste:2001gb,Cvetic:2001tj,Cvetic:2001nr,Blumenhagen:2002gw,Cvetic:2002pj,Honecker:2003vq,Cvetic:2003xs,Blumenhagen:2003jy,Cvetic:2004ui,Honecker:2004kb,Honecker:2004np,Blumenhagen:2004xx,Blumenhagen:2005tn,Gmeiner:2005vz,Bailin:2006zf,Bailin:2007va,Bailin:2008xx,Gmeiner:2008xq,Forste:2010gw,Bailin:2011am,Honecker:2012qr,Honecker:2013hda,Ecker:2014hma,Ecker:2015vea}. Some constructions have,
however, been directly performed on the type IIB side \cite{Bachas:1995ik,Blumenhagen:2000wh,Angelantonj:2000hi,Angelantonj:2000rw,Aldazabal:2000dg,Blumenhagen:2000ea,Marchesano:2004yq,Marchesano:2004xz,Dudas:2005jx,Kobayashi:2017dyu}.
Computing Yukawa couplings in the type IIB setting is useful also from an intersecting brane model builder's perspective. Type IIA Yukawa couplings have been computed in \cite{Cremades:2003qj}. There, they are given by sums over exponentials of classical worldsheet instanton actions. A factor in front of this sum cannot be fixed by classical
calculations. In \cite{Cremades:2004wa} also T-duality of intersecting brane models to type IIB flux compactifications is discussed. Couplings do match and further the type IIB calculation fixes the leading (in the small angle limit) contribution to the overall factor. Further discussions on the computation of interactions in type II models, including also quantum corrections, can be found in \cite{Aldazabal:2000cn,Abel:2002az,Cvetic:2003ch,Abel:2003fk,Abel:2003vv,Abel:2003yx,Lust:2004cx,Abel:2004ue,
Bertolini:2005qh,
Abel:2005qn,
Abel:2006yk,Duo:2007he,Russo:2007tc,
Billo:2007sw,Billo:2007py,
Pesando:2012cx,Pesando:2014owa,Pesando:2014sca,Pesando:2015fpj}.

Usually toroidal constructions are performed on so called factorisable six-tori consisting of three mutually orthogonal two-tori. Generalisations to non factorisable 
tori are studied in \cite{Blumenhagen:2004di,Faraggi:2006bs,Forste:2006wq,Forste:2007zb,Kimura:2007ey,Forste:2008ex,Bailin:2013sya,Forste:2014bfa,Berasaluce-Gonzalez:2016kqb,Berasaluce-Gonzalez:2017bib}. In particular in \cite{Forste:2014bfa} Yukawa couplings for
intersecting branes on non factorisable six-tori have been computed. The calculations are restricted to cases where the torus is generated by a sublattice of a lattice belonging to a factorisable torus; as a representative example the SO(12) root lattice is considered. In the present paper T-duality of this setup will be carried out. Yukawa couplings are found to match and the leading contribution to the overall factor can be computed in type IIB theory. Some technical details of the calculation are quite appealing. For instance, the SO(12) structure of the type IIA compactification is scrambled in the process of T-duality along some of the lattice vectors. It resurfaces at a later stage when zero modes of the Dirac equation are labelled. As an aside, the methods developed for $N_a$, $N_b$, $N_c$ not all beeing mutually coprime can easily
be applied to factorisable compactifications. In phenomenological model building such stiuations are not unlikely to arise; for instance an intitial $U(N)$ gauge symmetry can be broken by fluxes to Pati-Salam which in turn could be broken by Wilson lines to the Standard Model gauge group.

The paper is organised as follows. In the next section T-duality on the configuration of \cite{Forste:2014bfa} is performed. In section three, chiral fields as zero modes
of the Dirac equation are constructed. In section four, Yukawa couplings are computed by integrating the product of three zero modes over compact space. Section five contains some concluding remarks. In an appendix generalisations of the concept of greatest common divisors and lowest common multiples of lattices are reviewed and some examples given. 

\section{$\mathbf{D9}$ branes as T-dualised 
$\mathbf{D6}$ branes} 

In this section the T-dual of the setups considered in
\cite{Forste:2014bfa} will be constructed. The dual geometry will be a
six-torus whose complex structure matrix has off-diagonal
components. D-branes at angles give rise to magnetic flux, whereas
multiple intersections with the T-dualised cycle result in constant Wilson
lines.  
 
\subsection{T-dual of  $\mathbf{T^6_{\text{SO}(12)}}$: 
Closed String Sector}  

Before performing the T-duality, taking one from type IIA to type IIB,
the (deformed) six-torus on the type IIA side will be described
\cite{Forste:2014bfa}.  
 The compactification space is chosen to be a six dimensional flat
 torus $T^6$. It is given by the quotient space
 $\mathbb{R}^6/\Lambda^6$, where $\Lambda^6$ is a six dimensional
 lattice
\begin{equation*}
\Lambda^6=\Bigg\{\sum_{i=1}^6n_i\vec \alpha_i\Bigg|n_i\in\mathbb{Z}\Bigg\},
\end{equation*}
with $\{\vec \alpha_i\}_{i=1,...,6}$ generating the lattice. Hence,
locally the torus looks like $\mathbb{R}^6$, but points 
differing by lattice vectors are identified
\begin{equation*}
\vec x\sim \vec x+\vec \lambda,\quad x\in\mathbb{R}^6,\,\vec
\lambda\in\Lambda^6. 
\end{equation*}
In the following, the canonical basis of ${\mathbb R}^6$ will be
denoted by $\left\{ \vec e_i\right\}_{i=1,\ldots ,6}$ with components
\begin{equation}
e_{i\mu} = \delta_{i\mu} .
\label{eq:canbas}
\end{equation}
The metric on flat ${\mathbb R}^6$ is given by
\begin{equation*}
ds^2 =\sum_{h=1}^3 \left| d u_h\right| ^2 ,
\end{equation*}
where the six canonical coordinates have been combined into three complex
coordinates according to
\begin{equation}
\vec x = \sum_{i=1}^6 x_i \vec e_i = \sum_{h=1}^3
\text{Re}\left(u_h\right) \vec e_{2h-1} + 
\text{Im} \left( u_h\right) \vec e_{2h} .
\label{eq:compcon}
\end{equation} 
At the moment, this choice of pairs is arbitrary. Later D6 branes projecting 
onto straight lines in each of the
complex planes and thus automatically wrapping Lagrangian cycles
will be introduced. 

A torus is called factorisable if its generators $\left\{
  \vec\alpha_i\right\}$ can be split into three mutually
orthogonal pairs of  vectors. In this case, one would arrange the choice of
complex coordinates such that each of the mutually orthogonal pairs
lies within one complex plane. 
For non factorisable tori this is not possible. As a typical example the root
lattice of $SO(12)$,
\begin{eqnarray*}
\vec \alpha_1=(1,-1,0,0,0,0)^T,\quad \vec
  \alpha_2=(0,1,-1,0,0,0)^T,\quad \vec
  \alpha_3=(0,0,1,-1,0,0)^T,\\\nonumber 
\vec \alpha_4=(0,0,0,1,-1,0)^T,\quad \vec
  \alpha_5=(0,0,0,0,1,-1)^T,\quad \vec \alpha_6=(0,0,0,0,1,1)^T ,
\end{eqnarray*}
will be considered.
Here, vector components are given w.r.t.\ the canonical basis $\left\{
 \vec e_i\right\}$. If one was discussing just $T^6$ compactifications
without any further ingredients (such as D branes or envisaged
orientifolds) one could change metric and $B$ field components by arbitrary
constants. In particular, this allows deforming non factorisable into
factorisable tori. Here, additional ingredients allowing
deformations only within each of the complex planes will be assumed. 
This leads to the general metric 
\begin{equation*}
ds^2 = \sum_{h=1}^3 \frac{\text{Im} K_h}{\text{Im} \tau_h} \left| du_h\right|^2 ,
\end{equation*}
where $K_h$  and $\tau_h$ are complex parameters with positive
imaginary parts. The definition of 
the complex coordinates in (\ref{eq:compcon}) is also deformed
\begin{equation*}
u_h = x_{2h-1} + \tau_h x_{2h} .
\end{equation*}
In addition, a constant $B$ field of the form
\begin{equation*}
B = \sum_{h=1}^3 2\, \text{Re} K_h\, dx_{2h-1}\wedge dx_{2h} = \text{i}
\sum_{h=1}^3 \frac{\text{Re} K_h}{\text{Im}\tau_h} \, du_h \wedge d\bar{u}_h 
\end{equation*}
will be allowed.
For compactifications on a
factorisable $T^6$ the $K_h$'s would be the complexified
K{\"a}hler moduli of the three $T^2$'s whereas the $\tau_h$'s would
form the complex structure moduli. In \cite{Forste:2014bfa} it was
observed that Yukawa couplings of type IIA intersecting branes
exponentially depend
on these `would be' complex K{\"a}hler moduli even for non
factorisable $T^6$.  

Before performing T-duality, it is useful to change coordinates to the
lattice basis 
\begin{equation}
\sum_{i=1}^6 x_i \vec e_i = \sum_{i=1}^6 y_i \vec\alpha_i ,
\label{eq:ycord}
\end{equation}
such that integer shifts in any of the $y_i$ coordinates correspond to
lattice shifts.
Again, expressions for metric and $B$ field can be compressed by means
of complex coordinates
\begin{equation}
w_1 = y_1 +\frac{\tau_1\, y_2}{1-\tau_1}\,\,\, ,\,\,\,
w_2 = y_3 -\frac{y_2}{1-\tau_2} +\frac{\tau_2\, y_4}{1-\tau_2}\,\,\,
,\,\,\,
w_3 = y_5 -\frac{y_4}{1-\tau_3} +\frac{1 +\tau_3}{1-\tau_3}\, y_6 , 
\label{eq:comp2a}
\end{equation}
for which one obtains,
\begin{equation}
ds^2 =\sum_{h=1}^3 \frac{\text{Im} K_h \left|
    1-\tau_h\right|^2}{\text{Im}\tau_h} \left| dw_h \right|^2 \,\,\,
,\,\,\, B=\text{i} \sum_{h=1}^3 \frac{\text{Re} K_h \left|
    1-\tau_h\right|^2}{\text{Im}\tau_h}dw_h\wedge d\bar{w}_h .
\label{eq:b2a}
\end{equation}
Since the $y_i$ coordinates are compactified on circles they are
particularly useful for performing T-duality. 
The radii of these circles are taken to be at their selfdual value, 
$R= \sqrt{\alpha^\prime}$.
In the following 
\begin{equation}
\alpha^\prime = 1/\left(4\pi^2\right)
\label{eq:scalefix}
\end{equation}
will be chosen such that $2\pi R=1$. 
The Buscher rules  \cite{Buscher:1987sk} for
T-duality along the $\theta$ direction read
\begin{align}
\tilde{G}_{\theta\theta}  =  \frac{1}{G_{\theta\theta}} &,&
\tilde{G}_{ij} = G_{ij} -\frac{G_{\theta i}G_{\theta j} -B_{\theta
                                                            i}B_{\theta
                                                            j}}{G_{\theta\theta}}
& , &                                                          
\tilde{G}_{\theta i}  = \frac{B_{\theta
                               i}}{G_{\theta\theta}}
,\nonumber
  \\
\tilde{B}_{\theta i }  =  \frac{G_{\theta
                            i}}{G_{\theta\theta}} &,&
                            \tilde{B}_{ij}  = B_{ij} -\frac{ G_{\theta i}B_{\theta
                            j} -B_{\theta i}G_{\theta
                            j}}{G_{\theta\theta}} &, & 
\end{align}
where $i,j$ label directions other than $\theta$. In addition, there
is a shift in the dilaton 
\begin{equation}
    \Phi_b = \Phi_a - \frac{1}{2}\log G_{\vartheta\vartheta} .
    \label{eq:dilatonshift}
\end{equation}
The T-dual coordinate is again compactified on a circle of selfdual radius. 
Successively performing T-duality along the $y_1$, $y_3$ and $y_5$
direction yields type IIB theory. To write the T-dual background the following 
complex coordinates are introduced (omitting tildes at dual
coordinates)  
\begin{equation}
z_1 =y_1 + K_1\, y_2 \,\,\, ,\,\,\, z_2 = y_3 - K_2\, y_2
+ K_2\, y_4\,\,\, ,\,\,\, z_3 = y_5 - K_3\, y_4 + 2 K_3 \, y_6 .
\label{eq:compco}
\end{equation}
T-dual metric and $B$ field can be written as
\begin{align}
ds^2 & =  \sum_{h=1}^3 \frac{\text{Im}{\tau_h}}{\text{Im} K_h \left| 1
    -\tau_h\right|^2 }\left| dz_h\right|^2 ,\label{eq:metz}\\ 
B + dy_3\wedge dy_4 -2 dy_5 \wedge dy_6 & = 
\frac{\text{i}}{\text{Im}K_1}\text{Re}\frac{\tau_1}{1-\tau_1}
dz_1\wedge d\bar{z}_1 + \frac{\text{i}}{\text{Im}K_2} \text{Re}
                                          \frac{1}{1-\tau_2}  
dz_2 \wedge d\bar{z}_2\nonumber \\ &   +\frac{\text{i}}{\text{Im}
K_3}\text{Re}\frac{1}{1- \tau_3} dz_3\wedge d\bar{z}_3 .
\end{align}
Here, moduli have been suggestively split into complex structure
appearing in (\ref{eq:compco}) and the rest. This is not unique. The
6d metric has 21 independent real components whereas complex structure
moduli and imaginary part of the K{\"a}hler moduli have 18 plus 9 real
components. Uniqueness is achieved by imposing the six additional
conditions that the $B$ field should have components only along
$(1,1)$ forms \cite{Moore:1998pn}\footnote{The same
  argument can be also applied to the type IIA side. From (\ref{eq:comp2a}) and
  (\ref{eq:b2a}) one learns that actual complex structure moduli are
  given purely in terms of `would be' complex structure moduli,
  independent of `would be' K{\"a}hler moduli.}. 
To achieve that, the complex structure will not be modified but instead T-duality will
be combined with the gauge transformation 
\begin{equation}
B \to  B -2 dy_3\wedge dy_4 -2 dy_5 \wedge dy_6 , 
\label{eq:gaugetr}
\end{equation}                                 
which has to be kept in mind when performing T-duality in the
open string sector. Notice, that the previously `would be' K{\"a}hler
moduli become actual complex structure moduli in the T-dual type IIB
theory. Finally, the relation between type IIB and type IIA dilaton is
\begin{equation}
    \Phi_b =\Phi_a -\frac{1}{2}\sum_{h=1}^3 \log\frac{\text{Im} K_h\left| 1-\tau_h\right|^2}{\text{Im}\tau_h} .
    \label{eq:expdils}
\end{equation}

\subsection{T-dual of  $\mathbf{T^6_{\text{SO}(12)}}$: 
Open String Sector}  

As discussed in \cite{Forste:2007zb,Forste:2014bfa} a D6 brane of
type IIA theory spans the following three dimensional subspace of the
six dimensional compact space
\begin{equation}
x_{2h} = \frac{m^h}{n^h} x_{2h -1},\,\,\, \text{for}\,\,\, h \in
\left\{ 1,2,3\right\} .
\label{eq:dbrane}
\end{equation}
For factorisable tori the wrapping numbers $n^h$ and $m^h$ should be
coprime for each $h$. In the non factorisable case these conditions
are modified. For instance if the pairs are still all coprime, $n^h +
m^h$ has to be even for all $h$'s. Other possibilities are listed in
\cite{Forste:2007zb,Forste:2014bfa}. For simplicity, the
case that branes pass through the origin will be considered. If one of 
the wrapping
numbers $n^h$ is zero the corresponding equation has to be replaced by
$x_{2h-1} =0$.  Expressed in $y_i$ coordinates (\ref{eq:ycord}),
  equations (\ref{eq:dbrane}) take the form
\begin{align}
& y_1 = \frac{ n^1 \, y_2}{N^{(1)}},\,\,\, y_3 = \frac{m^2 \, y_2 +
  n^2 y_4}{N^{(2)}},\,\,\, y_5 = \frac{m^3\, y_4 +\left( n^3 -
    m^3\right) y_6}{N^{(3)}} ,\nonumber\\ & \text{with}\,\,\, N^{(h)} = n^h +
m^h,\,\,\, h\in \left\{ 1,2,3\right\} 
\label{eq:dcond}
\end{align}
In the following, the case that any of the $N^{(h)}$
vanishes will be excluded, i.e.\ T-duality 
along a D-brane will not be performed. 
This case has to be treated separately and leads to D7, D5 or D3 branes in the
T-dual picture. T duality for open strings has been discussed in e.g.\ in
\cite{Dai:1989ua,Horava:1989ga,Green:1991et,Alvarez:1996up,Dorn:1996an,Forste:1996hy,Borlaf:1996na}. Eq.\ (\ref{eq:dcond}) represents Dirichlet conditions on the coordinates
with respect to  which T-duality will be performed. Dirichlet conditions turn 
into Neumann
conditions, which are obtained by varying the worldsheet action with
no boundary conditions on the variation and a gauge field coupling to
the boundary. This gauge field is given by minus the right hand sides of
(\ref{eq:dcond}),
\begin{equation*}
\frac{\tilde{A}_{y_1}}{2\pi} = -\frac{ n^1 \, y_2}{N^{(1)}},\,\,\,
\frac{\tilde{A}_{y_3}}{2\pi} = -\frac{m^2 \, y_2 + 
  n^2 y_4}{N^{(2)}},\,\,\, \frac{\tilde{A}_{y_5}}{2\pi} = -\frac{m^3\, y_4
  +\left( n^3 - m^3\right) y_6}{N^{(3)}} .
\end{equation*}
As to be discussed shortly, these gauge fields are multiplied by identity 
matrices whose appearance has been supressed so far. 
The T-dual fieldstrength is finally computed as (recall
(\ref{eq:gaugetr}))
\begin{align}
F & = \frac{1}{2} F_{ij}\, dy_i\wedge dy_j = d\tilde{A} +\tilde{A}\wedge \tilde{A}
-2\pi \left( dy_3\wedge
    dy_4 + dy_5\wedge 
    dy_6\right)\nonumber\\
& = \frac{\text{i}\pi}{\text{Im} K_1}\frac{ n^1}{N^{(1)}}dz_1\wedge
  d\bar{z}_1 - \frac{\text{i}\pi}{\text{Im} K_2} 
  \frac{m^2}{N^{(2)}} dz_2\wedge d\bar{z}_2 -
  \frac{\text{i}\pi}{\text{Im} K_3}\frac{m^3}{N^{(3)}} dz_3\wedge
  d\bar{z}_3,
\end{align} 
where T-dual complex coordinates are defined in (\ref{eq:compco}). 
It is consistent that starting with D branes wrapping Lagrangian
cycles in type IIA theory the T-dual D9 branes of type IIB carry flux
only along (1,1) forms.
So far, multiple wrappings of the D9 brane have not been taken into account.
The D9 brane wrapping
number, $N=N_{\text{D9}}$, is given by
\begin{equation*}
N=N_{\text{D9}} = \frac{N^{(1)}N^{(2)}N^{(3)}}{2} N_{\text{D6}} ,
\end{equation*}
where $N_{\text{D6}}$ is the wrapping number of the D6 brane and the
additional multiplicity originates from the intersection number with
the cycle along which T-duality has been performed. (Intersection
numbers for the type IIA setting are taken from
\cite{Forste:2007zb}.) In the following 
$$  
N_{\text{D6}} = 1
$$
will be considered
since for the calculation of Yukawa couplings this number is not
relevant. (Given a gauge group $U\left( A \right) \times U\left(
  B\right) \times U\left( C\right)$ the Yukawa coupling of $\left(
  A\overline{B}\right) \left( B\overline{C}\right) \left( C
  \overline{A}\right)$ does not depend on $A$, $B$, $C$.) 
For later convenience, the gauge transformation
(\ref{eq:gaugetr}) will be included in a redefinition of the T-dual gauge field. 
Taking into account multiple wrappings, 
\begin{equation}
F = dA + A \wedge A\,\,\, ,\,\,\, A =   \left(\frac{\pi n^1\text{Im}\bar{z}_1
  dz_1}{N^{(1)}\text{Im} K_1}
+ \frac{\pi m^2 \text{Im} z_2 d\bar{z}_2}{N^{(2)}\text{Im} K_2} +
  \frac{\pi m^3\text{Im} z_3 d\bar{z}_3}{N^{(3)}\text{Im} K_3}\right)
\mathbbm{1}_N + W.
\label{eq:fluwi}
\end{equation}  
is chosen, where $W$ is a Wilson line originating
from the finite separation of $N^{(1)}N^{(2)}N^{(3)}/2$ stacks of
branes along the T-dualised direction. Although $W$ can be written as
$g^{-1} dg$ with $g \in SU(N)$ it cannot be removed by a globally
single valued gauge transformation. On the type IIB side, it breaks
the gauge group from $U\left(N\right)$ to
$U\left(N_{\text{D6}}=1\right)$. The Wilson line will be discussed more
explicitly in the next section.
 
\section{Chiral Matter} 

This section follows closely the strategy of
\cite{Cremades:2004wa} in identifying chiral matter of the effective
four dimensional theory. First, Wilson lines are specified.  They are
viewed as gauge transformations induced by lattice shifts. In the
factorisable case these gauge transformations are associated to the direct
product of three matrices, or in other words, each of the two group
indices on the gauge transformation matrix is conveniently replaced by a
triplet of indices. It will be argued that in the non factorisable case
the gauge index should be expressed in terms of a vector in a quotient
lattice. To really discuss the T-dual of intersecting branes, more than
one unitary gauge group factor has to be considered. Zero modes of the 
Dirac equation in the bifundamental representation will give rise to
chiral matter.  

\subsection{Labelling Gauge Indices \label{sec:labelling}}

Consider a field $\phi $ as a function of torus
coordinates transforming in the fundamental representation of
$U(N)$. Dependence on uncompactified spacetime is also assumed 
but suppressed in the notation. It is imposed that this field is
periodic under lattice shifts up to gauge transformations, i.e.\
\begin{equation}
\phi\left( y_1,\ldots, y_i + 1,\ldots, y_6\right) =
\text{e}^{\text{i}\chi_i\left( \vec{z}\right)} \omega_i \,  \phi\left(
  y_1,\ldots, y_i ,\ldots, y_6\right) ,
\label{eq:embed}
\end{equation}
where $\chi_i$ contains effects due to magnetic flux (\ref{eq:fluwi}),
\begin{equation}
\chi_i \left( \vec{z}\right) = \oint_{\left(
  y_1,\ldots, y_i ,\ldots, y_6\right)} ^{ \left( y_1,\ldots, y_i +
  1,\ldots, y_6\right)} \left( A - W\right) .
\end{equation}
The Wilsonline $W$ has been encoded in a constant matrix $\omega_i \in
SU(N)$. The phases $\chi_i$ are explicitly given by 
\begin{align}\nonumber
&\chi_1\left(\vec z\right) = -\frac{\pi n^1}{N^{(1)}}
  \frac{\text{Im}\left(z_1\right)}{\text{Im}(K_1)}, \quad\chi_2\left(\vec
  z\right)= -\frac{\pi n^1}{N^{(1)}}\frac{\text{Im}\left(\overline{K_1}
  z_1\right)}{\text{Im}\left(K_1\right)}+\frac{\pi
  m^2}{N^{(2)}}\frac{\text{Im}\left(\overline{K_2}
  z_2\right)}{\mathrm{Im}(K_2)},\\ \label{eq:phases} &\chi_3\left(\vec
  z\right)=-\frac{\pi 
  m^2}{N^{(2)}}\frac{\text{Im}\left(z_2\right)}{\text{Im}\left( K_2\right)},
  \quad  
\chi_4\left(\vec z\right)=-\frac{\pi
  m^2}{N^{(2)}}\frac{\text{Im}\left(\overline{K_2}
  z_2\right)}{\text{Im}\left( K_2\right)} + \frac{\pi
  m^3}{N^{(3)}}\frac{\text{Im}\left(\overline{K_3}
  z_3\right)}{\text{Im}\left( K_3\right)},\\ \nonumber &\chi_5(\vec
                                                         z)=-\frac{\pi 
  m^3}{N^{(3)}}\frac{\mathrm{Im}(z_3)}{\text{Im}\left( K_3\right) },\quad
  \chi_6\left(\vec z\right)=-\frac{2\pi
  m^3}{N^{(3)}}\frac{\text{Im}\left( \overline{K_3}z_3\right)}
  {\text{Im}\left( K_3\right)}.   
\end{align}
The $SU(N)$ factors $\omega_i$ will be fixed by consistency. Taking
the argument once through a closed loop should leave a field
transforming in the fundamental representation invariant, i.e.\
\begin{equation}\label{condi}
{\omega_j}^{-1}{\omega_i}^{-1}\omega_j\omega_i \phi\left(\vec
z\right)=\mathrm{e}^{2\pi i k_{ij}/N}\cdot\mathbbm{1}_N\cdot\phi\left(\vec
z\right) .
\end{equation}
The phases are fixed such that a phase originating from $A-W$ is
cancelled, e.g.\
\begin{equation*}
k_{12} = - \frac{N}{2\pi}
\left( \chi_1\left( 0,0,0 \right) + \chi_2\left(
    1,0,0\right) - \chi_1\left( 1 +K_1, -K_2,0\right) - \chi_2\left(
    K_1, -K_2,0\right)\right)\,\,\, \text{mod}\,\,\, N
\end{equation*}
Non vanishing phases are
\begin{align}\nonumber
& k_{12}=\frac{n^1}{2}N^{(2)}N^{(3)} \mod N,\quad
  k_{32}=\frac{m^2}{2}N^{(1)}N^{(3)} \mod N,\\ \nonumber
&  k_{34}=-\frac{m^2}{2}N^{(1)}N^{(3)}\mod N,\quad 
 k_{54}=\frac{m^3}{2}N^{(1)}N^{(2)}\mod N,\\
&  k_{56}=-m^3N^{(1)}N^{(2)}\mod N .
\label{k}
\end{align}
Notice, that all $k_{ij}$'s are integers. This is related to
conditions that D6 branes should wrap closed cycles in the type IIA
geometry \cite{Forste:2014bfa,Forste:2007zb}.  
One may try to construct the $\omega_i$'s by means of two
dimensional solutions given in \cite{Cremades:2004wa}: Consider two
matrices $w_1,w_2\in SU(n)$, where $n\in\mathbb{Z}_+$. Impose the condition
\begin{equation*}
{w_2}^{-1}{w_1}^{-1}w_2 w_1=\mathrm{e}^{2\pi i k/n}\cdot \mathbbm{1}_\text{n} .
\end{equation*}
A solution for $w_1$
and $w_2$ is
\begin{equation*}
w_1=Q^m,\quad w_2=P
\end{equation*}
where $m=k\mod n$ and
\begin{equation*} 
Q=\begin{pmatrix}1&&&\\&\mathrm{e}^{2\pi
    i/n}&&\\&&\ddots&\\&&&\mathrm{e}^{2\pi
    i(n-1)/n}\end{pmatrix},\quad\text{and}\quad
P=\begin{pmatrix}&1&&\\&&\ddots&\\&&&1\\1&&&\end{pmatrix}. 
\end{equation*}
For the factorisable torus solutions of the form
\begin{align}\nonumber
&\omega_1=Q_{(1)}^{n^1}\otimes\mathbbm{1}_{(2)}\otimes\mathbbm{1}_{(3)},\quad
  \omega_3=\mathbbm{1}_{(1)}\otimes
  Q^{-m^2}\otimes\mathbbm{1}_{(3)},\quad
  \omega_5=\mathbbm{1}_{(1)}\otimes\mathbbm{1}_{(2)}\otimes
  Q_{(3)}^{-m^3},\\\label{sol} 
&\omega_2=P_{(1)}\otimes P^{-1}_{(2)}\otimes
  \mathbbm{1}_{(3)},\quad\omega_4=\mathbbm{1}_{(1)}\otimes
  P_{(2)}\otimes
  P^{-1}_{(3)},\quad\omega_6=\mathbbm{1}_{(1)}\otimes\mathbbm{1}_{(2)}\otimes
  P^{2}_{(3)}, 
\end{align}
where matrices with subscript $(h)$ are $N^{(h)}\times N^{(h)}$
matrices, solve conditions corresponding to (\ref{condi})
\cite{Cremades:2004wa}. The resulting $\omega_i$ are $2N\times2N$
matrices. A similar overcounting arises on the type IIA side if one
just multiplied the intersections numbers in each complex plane
\cite{Forste:2014bfa}. There the overcounting would happen due to an
erroneous labelling of intersection points as $j^{(h)} \in {\mathbb
  Z}_{N^{(h)}}$. The resolution advocated in \cite{Forste:2014bfa} is
that the triplet of $j^{(l)}$'s takes values in a sublattice of $
  \prod_{h=1}^3 {\mathbb
  Z}_{N^{(h)}}$. 
With triple indices $i,j$, (\ref{sol}) reads
\begin{align} \nonumber
&(\omega_1)_{ij} =
  Q^{n^1}_{i^{(1)},j^{(1)}}\delta_{i^{(2)},j^{(2)}}\delta_{i^{(3)},j^{(3)}},\,\,\,
\left(\omega_2\right)_{ij}=P_{i^{(1)},j^{(1)}}P^{-1}_{i^{(2)},j^{(2)}}
  \delta_{i^{(3)},j^{(3)}},
 \\ \label{eq:solcomp} 
& \left(\omega_3\right)_{ij} = \delta_{i^{(1)},j^{(1)}}Q^{-m^2}_{i^{(2)},j^{(2)}}
  \delta_{i^{(3)},j^{(3)}},\,\,\,
\left(\omega_4\right)_{ij} = \delta_{i^{(1)}j^{(1)}}P_{i^{(2)}j^{(2)}}
  P^{-1}_{i^{(3)}j^{(3)}},  
  \\\nonumber      
& \left(\omega_5\right)_{ij} = \delta_{i^{(1)},j^{(1)}}
  \delta_{i^{(2)},j^{(2)}}Q^{-m^3}_{i^{(3)},j^{(3)}},\,\,\,
\left(\omega_6\right)_{ij} = \delta_{i^{(1)}j^{(1)}}\delta_{i^{(2)}j^{(2)}}
  P^{2}_{i^{(3)}j^{(3)}}  .
\end{align}
It remains to identify the lattice $\Lambda^3$ within which triple
indices take values.  
Wrapping numbers on the type IIA side describe closed cycles if one of
the following four cases applies: 
\begin{itemize}
\item[$(i)$]  all three $N^{(h)}$'s are even and all $\left( n^h,
    m^h\right)$ are coprime,
\item[$(ii)$]  all three $N^{(h)}$'s are even but for exactly one $i$:
  $g.c.d. \left( n^i, m^i\right) =2$, remaining are coprime pairs,
\item[$(iii)$] two $N^{(h)}$'s are even, for exactly one $i$:
  $g.c.d. \left( n^i, m^i\right) =2$, remaining are coprime pairs,
\item[$(iv)$] one $N^{(h)}$ is even and for the corresponding pair
  $g.c.d. \left( n^h, m^h\right) =2$, remaining are coprime pairs .
\end{itemize}
Whenever $g.c.d.\left( n^i, m^i\right) =2$, $N^{(i)}/2$ has to be
odd. Otherwise the corresponding wrapping numbers have to be divided
by two and another case applies. 
The following quotient lattices, $\Lambda^3$, turn out to yield useful
sets for labels:
\begin{itemize}
\label{list:indexlat}
\item[$(i)$, $(ii)$] $\Lambda_3 = \Lambda_{\text{SO(6)}}/
  \bigotimes_{l=1}^3 N^{(l)}{\mathbb Z}$
\item[$(iii)$] $\Lambda_3 = \Lambda_{\text{SO(6)}}/\Gamma$. If e.g.\
  $g.c.d.\left( n^1, m^1\right) = 2$ and $N^{(2)}$ odd then $\Gamma$
is generated by $\left( N^{(1)}/2, N^{(2)},0\right)$, $\left( N^{(1)}/2,
 -N^{(2)},0\right)$, $\left( 0,0,N^{(3)}\right)$
\item[$(iv)$]  $\Lambda_3 = \Lambda_{\text{SO(6)}}/\Gamma$. If e.g.\
  $g.c.d.\left( n^1, m^1\right) = 2$ then $\Gamma$ is generated by
  $\left( N^{(1)}/2, N^{(2)},0\right)$, $\left( N^{(1)}/2, 
 -N^{(2)},0\right)$, $\left( 0,N^{(2)},N^{(3)}\right)$.
\end{itemize}
Here, $\Lambda_{\text{SO(6)}}$ denotes the $SO(6)$ root lattice
generated by $\left( 1,-1,0\right)$, $\left(0,1,1\right)$, $\left(
  0,1,-1\right)$. 
In cases where $g.c.d.\left( n^h, m^h\right) = 2$ the corresponding $Q$
matrix has to be replaced by 
\begin{equation*}
Q_{N^{(h)}}\to Q_{N^{(h)}/2}\otimes \mathbbm{1}_{2}\,\,\,
  \text{if}\,\,\, g.c.d.\left( n^h ,m^h\right) =2 .  
\end{equation*}
\subsection{Bifundamentals}
Consider two D6 branes wrapping cycles labelled by $a$ and $b$ on the
type IIA side. Assume that neither cycle has zero intersection number
with the T-dualised cycle. On the type IIB side this corresponds to
$N_a + N_b$ D9 branes wrapping the T-dual six torus where $N_a$ and
$N_b$ are the respective intersection numbers with the T-dualised
cycle. These D9 branes accommodate a $U\left( N_a + N_b\right)$ gauge
symmetry which is broken to $U\left( N_a\right) \times U\left(
  N_b\right)$ by magnetic fluxes and finally to $U\left( 1\right)
\times U\left( 1\right)$ by Wilson lines. The magnetic flux is given
by the following non vanishing fieldstrength components
\begin{align}
& F_{z_1\overline z_1}=\frac{\pi
  i}{\mathrm{Im}\left(K_1\right)} \begin{pmatrix}
  \frac{n^1_a}{N^{(1)}_a}\mathbbm{1}_{N_a}   
  \\
& \frac{n^1_b}{N^{(1)}_b}\mathbbm{1}_{N_b}\end{pmatrix}, 
\quad F_{z_2\overline z_2}=-\frac{\pi
  i}{\mathrm{Im}\left(
  K_2\right)}\begin{pmatrix}\frac{m^2_a}{N^{(2)}_a}\mathbbm{1}_{N_a}& 
\label{bifundamentalfieldstrength}  
  \\ &\frac{m^2_b}{N^{(2)}_b}\mathbbm{1}_{N_b}\end{pmatrix},\\\nonumber
& F_{z_3\overline z_3}=-\frac{\pi
  i}{\mathrm{Im}\left
  (K_3\right)}\begin{pmatrix}\frac{m^3_a}{N^{(3)}_a}\mathbbm{1}_{N_a} 
  &\\&\frac{m^3_b}{N^{(3)}_b}\mathbbm{1}_{N_b}\end{pmatrix}.
\end{align}
Let $\phi$ be a field transforming in the $\left( N_a,
  \overline{N_b}\right)$ representation of $U\left( N_a\right) \times
U\left( N_b\right)$. Formula (\ref{eq:embed}) is modified to
\begin{equation}
\phi\left( y_1,\ldots, y_i + 1,\ldots, y_6\right) =
\text{e}^{\text{i}\chi_i^{ab}\left( \vec{z}\right)}\,
\omega_i ^a \phi\left(
  y_1,\ldots, y_i ,\ldots, y_6\right){\omega_i^b}^\dagger ,
\label{eq:biembed}
\end{equation}
where $\chi_i^{ab} = \chi_i^a -\chi_i^b$ denotes the difference
between the two phases (\ref{eq:phases}). Defining
\begin{equation*}
I_{ab}^{(h)} = n_a^h m_b^h - m_a^h n_b^h\,\,\, ,\,\,\, \tilde{I}_{ab}^{(h)}
=  I_{ab}^{(h)}/N_a^{(h)} N_b^{(h)} ,
\end{equation*}
the phase differences can be written as
\begin{align}\nonumber
&\chi_1^{ab}\left(\vec z\right) =
  \frac{\pi\text{Im}\left(z_1\right)\tilde{I}^{(1)}_{ab}}{\text{Im}(K_1)},
  \quad\chi_2^{ab}\left(\vec 
  z\right)= \frac{\pi\text{Im}\left(\overline{K_1}
  z_1\right)\tilde{I}^{(1)}_{ab}}{\text{Im}\left(K_1\right)}
  -\frac{\pi \text{Im}\left(\overline{K_2}
  z_2\right)\tilde{I}^{(2)}_{ab}}{\mathrm{Im}(K_2)},\\ \label{eq:biphases}
&\chi_3^{ab}\left(\vec 
  z\right)=\frac{\pi\text{Im}\left(z_2\right)
  \tilde{I}^{(2)}_{ab}}{\text{Im}\left( 
  K_2\right)},
  \quad  
\chi_4^{ab}\left(\vec z\right)=\frac{\pi\text{Im}\left(\overline{K_2}
  z_2\right)\tilde{I}^{(2)}_{ab}}{\text{Im}\left( K_2\right)} -
  \frac{\pi\text{Im}\left(\overline{K_3} 
  z_3\right)\tilde{I}^{(3)}_{ab}}{\text{Im}\left( K_3\right)},\\
  \nonumber 
&\chi_5^{ab}(\vec z) = \frac{\pi\mathrm{Im}(z_3)
  \tilde{I}^{(3)}_{ab}}{\text{Im}\left( K_3\right) },\quad 
  \chi_6^{ab}\left(\vec z\right)=\frac{2\pi \text{Im}\left(
  \overline{K_3}z_3\right) \tilde{I}^{(3)}_{ab}} 
  {\text{Im}\left( K_3\right)}.   
\end{align}
Inserting the explicit representations for the Wilson lines
(\ref{eq:solcomp}) one finds
\begin{align}\nonumber
\left(\omega_1^a\phi(\vec z)\omega_1^{b\dag}\right)_{k_ak_b} 
& = \text{e}^{2\pi \text{i} \left(
      k_a^{(1)}n_a^1/N_a^{(1)}-k_b^{(1)}n_b^1/N_b^{(1)} \right)}
      \phi_{k_ak_b}(\vec z),\\
\nonumber  \left(\omega_2^a\phi \left(\vec z\right)\omega_2^{b\dag}\right)_{k_ak_b}
& =\phi_{k_a+(1,-1,0),k_b+(1,-1,0)}(\vec z), \\\nonumber
\left(\omega_3^a\phi\left(\vec z\right)\omega_3^{b\dag}\right)_{k_ak_b}
& = \text{e}^{-2\pi \text{i}
  \left(k_a^{(2)}m_a^2/N_a^{(2)}-k_b^{(2)}m_b^2/N_b^{(2)}\right)}
  \phi_{k_ak_b}\left(\vec z\right),\\ \label{phicomptrafo}
\left(\omega_4^a\phi_{k_ak_b}\left(\vec z\right) \omega_4^{b\dag}\right)_{k_ak_b}
& = \phi_{k_a+(0,1,-1),k_b+(0,1,-1)}\left(\vec z\right) ,\\ \nonumber 
\left(\omega_5^a\phi\left(\vec z\right)\omega_5^{b\dag}\right)_{k_ak_b} 
& =  \text{e}^{-2\pi \text{i}
  \left(k_a^{(3)}m_a^3/N_a^{(3)}-k_b^{(3)}m_b^3/N_b^{(3)}\right)}
  \phi_{k_ak_b}\left(\vec z\right), \\\nonumber  
\left(\omega_6^a\phi\left(\vec z\right)\omega_6^{b\dag}\right)_{k_ak_b}
& = \phi_{k_a+(0,0,2),k_b + (0,0,2)}\left(\vec z\right).
\end{align}
Notice that $\omega_2$, $\omega_4$ and $\omega_6$ act as shifts by
$SO(6)$ roots on the gauge indices $k_a$ and $k_b$. Therefore the
convention to label the gauge group elements by a subset of $SO(6)$
roots is consistent with gauge transformations. In
\cite{Cremades:2004wa} it is demonstrated that replacing the double
index at matrix components by a single index is very useful. The
details can be summarised as follows. Focusing on just one $T^2$
factor the expression corresponding to the first line in
(\ref{phicomptrafo}) reads ($\varphi$ replaces $\phi$ for the case of
two extra dimensions)
\begin{equation}
\left( \omega^a \varphi\left( z\right) \omega^{b\dag}\right)_{k_a k_b} =
\text{e}^{2\pi \text{i} \left( \frac{k_a n_a}{N_a} -\frac{k_b
      n_b}{N_b}\right)} \varphi_{k_ak_b}\left( z\right) =
\text{e}^{2\pi \text{i} \tilde{I}_{ab} \ell }\varphi_{\ell \, \ell} ,
\label{eq:introl}
\end{equation}
where in the last step the double index has been replaced by a single
index
\begin{equation}
\ell \in \left\{ 0, \ldots , N_a N_b -1\right\},
\label{eq:lrange}
\end{equation}
from which it is obtained by
\begin{equation}
k_a =
\ell\mod N_a\,\,\, ,\,\,\, k_b = \ell \mod N_b .
\label{eq:doubfs}
\end{equation}   
This means that there is a pair of integers $\left( s,t\right)$ such
that
\begin{equation*}
\ell = k_a + s N_a = k_b + tN_b  \mod N_a N_b ,
\end{equation*}
implying that the difference $k_a - k_b$ has to be an integer multiple
of
$$d= g.c.d.\left( N_a , N_b\right) . $$
For this reason the discussion in \cite{Cremades:2004wa} is restricted
to the case $d=1$. For general $d$, the intersection number 
$I_{ab} = n_a N_a - n_b N_b$ is a multiple of $d$. Hence,
(\ref{eq:lrange}) should be raplaced by 
\begin{equation}
\ell \in \left\{ 0, \ldots , \frac{N_a N_b}{d} -1\right\},
\label{eq:modlrange}
\end{equation}
providing not enough labels. In addition, one should introduce another
label
\begin{equation*}
\delta \in \left\{ 0,\ldots, d-1 \right\} ,
\end{equation*}
with
\begin{equation}
k_a - k_b = 0 \mod d \,\,\, \longrightarrow\,\,\, k_a - k_b = \delta \mod d.
\label{eq:kdiff}
\end{equation}
The distribution of $\delta$ among individual shifts of $k_a$ and
$k_b$ is carried out as follows. First, one chooses a solution $\left(
  p,q\right)$ of the linear Diophantine equation
\begin{equation*}
d = N_a p - N_b q .
\end{equation*}
Then (\ref{eq:kdiff}) is compatible with
\begin{equation*}
k_a \to k_a + \frac{N_a p \delta}{d}\mod N_a \,\,\, ,\,\,\, k_b \to k_b+
\frac{N_bq\delta}{d} \mod N_b .
\end{equation*}
Summarising, the correspondence between $\left( k_a , k_b\right)$ and
$\left( \ell , \delta\right)$ is
\begin{equation}
k_a = \ell + \frac{p N_a \delta}{d} \mod N_a \,\,\, ,\,\,\, k_b = \ell
 +  \frac{q N_ab\delta}{d} \mod N_b . 
\label{eq:soofnice}
\end{equation}
Then, the second identity in (\ref{eq:introl}) generalises to
\begin{equation}
\text{e}^{2\pi \text{i} \left( \frac{k_a n_a}{N_a} -\frac{k_b
      n_b}{N_b}\right)} \varphi_{k_ak_b}\left( z\right) =
\text{e}^{2\pi \text{i} \left(\tilde{I}_{ab} \ell + \frac{
      \delta}{d}\left( p n_a + q n_b\right)\right)}\varphi_{\ell +
  \frac{pN_a\delta}{d}\, \ell + \frac{qN_b\delta}{d}} ,  
\label{eq:phasfac}
\end{equation} 
i.e.\ there is an additional phase taking values in ${\mathbb Z}_d$.

Returning to the non factorisable torus, discussed previously, it was
noticed that
\begin{equation*}
k_a \in \frac{\Lambda_{\text{SO(6)}}}{\Gamma_a}\,\,\, ,\,\,\, k_b \in
\frac{\Lambda_{\text{SO(6)}}}{\Gamma_b} , 
\end{equation*}
where $\Gamma_a$, $\Gamma_b$ are sublattices of
$\Lambda_{\text{SO(6)}}$. 

The lattice $\Gamma_d$ is defined as
follows: $\Gamma_a$ and $\Gamma_b$ are sublattices of $\Gamma_d$ and
there is no proper sublattice of $\Gamma_d$ containing $\Gamma_a$ and
$\Gamma_b$ as sublattices. In other words, $\Gamma_d$ is the coarsest
lattice containing $\Gamma_a$ and $\Gamma_b$. 

The number of inequivalent index combinations $\left( k_a,
  k_b\right)$ is given in terms of indices of quotient
lattices\footnote{The index of a quotient lattice counts how often the
  fundamental cell of the lattice fits into the fundamental cell of
  the sublattice with respect to which the quotient is taken.}
\begin{equation}
\#\left( k_a, k_b\right) =
\left|\frac{\Lambda_{\text{SO(6)}}}{\Gamma_a}\right|
\left|\frac{\Lambda_{\text{SO(6)}}}{\Gamma_b}\right|  = 
\left| \frac{\Lambda_{\text{SO(6)}}}{\Gamma_a \cap \Gamma_b}\right|
  \left| \frac{\Lambda_{\text{SO(6)}}}{\Gamma_d}\right| .
\label{eq:indexrange}
\end{equation}
The second equality with a reference to its proof is discussed further
in appendix 
\ref{ap:proof}. These observations suggest replacing the index pair
$\left( k_a, k_b\right)$ by two lattice valued labels
\begin{equation*}
l \in \frac{\Lambda_{\text{SO(6)}}}{\Gamma_a \cap \Gamma_b}\,\,\,
,\,\,\, \delta \in \frac{\Lambda_{\text{SO(6)}}}{\Gamma_d} .
\end{equation*}
The pair $\left( k_a , k_b\right)$ can be again obtained by shifting
values of $l$ and modding out by lattices $\Gamma_a$, respectively
$\Gamma_b$. The details are as follows. There are classes of
differences $k_a - k_b$ labelled by different $\delta$'s, 
\begin{equation}
k_a - k_b = \delta .
\label{eq:delrel}
\end{equation}
Throughout the paper three dimensional lattice vectors are viewed as a
column with three entries corresponding to the components with respect
to a given basis (mostly (\ref{eq:canbas})). Let $a_i$, $b_i$, $d_i$
($i\in \left\{ 1,2,3\right\}$) be the generators of the lattices
$\Gamma_a$, $\Gamma_b$, $\Gamma_d$, respectively. It turns out to be
convinient to combine these into three by three matrices
\begin{equation}
A =\left( a_1, a_2, a_3\right)\,\,\, ,\,\,\, B=\left( b_1, b_2,
  b_3\right)\,\,\, ,\,\,\, D=\left( d_1, d_2 , d_3\right) .
\label{eq:latmat}
\end{equation}
The requirement that $\Gamma_a$ and $\Gamma_b$ are sublattices of
$\Gamma_d$ is equivalent to the existence of three by three integral
matrices\footnote{Matrices with integer components are called integral
  matrices.} 
$M_a$ and $M_b$ such that
\begin{equation}
A = D M_a \,\,\, ,\,\,\, B= D M_b .
\label{eq:latemb}
\end{equation}
Let $\left( P, Q\right)$ be two three by three integral matrices
satisfying\footnote{These matrices exist, reference to a 
  proof is given in appendix \ref{ap:proof}.}
\begin{equation}
D = A P  - B Q.
\label{eq:DAB}
\end{equation}
A natural generalisation of (\ref{eq:soofnice}) would be 
\begin{equation*}
k_a  = l+A P D^{-1}\delta\mod \Gamma_a \,\,\, ,\,\,\, k_b =
l + B QD^{-1}\delta  \mod \Gamma_b  .
\end{equation*}
There is however a problem with that. The partitions $A P D^{-1}\delta$ and 
$B QD^{-1}\delta$ are not always in $\Lambda_{\text{SO(6)}}$. A way out
is to give up invariance under equivalence shifts of $\delta$. So,
in the following $\delta$ will be taken from a finite set consisting
of one representative for each equivalence class. Then it makes sence
to assign
\begin{equation*}
k_a  = l\mod \Gamma_a \,\,\, ,\,\,\, k_b =
l -\delta  \mod \Gamma_b  ,
\end{equation*}
since now shifts of $k_b$ by elements of $\Gamma_d$ cannot be absorbed
by picking another $\delta$ from the same equivalence class. 
The non factorisable version of (\ref{eq:phasfac}) reads
\begin{equation*}
\text{e}^{2\pi \text{i} \left( \frac{k_a^{(1)} n^1_a}{N_a^{(1)}}
    -\frac{k_b^{(1)} 
      n_b^1}{N_b^{(1)}}\right)} \phi_{k_ak_b} =
\text{e}^{2\pi \text{i} \left(\tilde{I}_{ab} l^1                 
    +\frac{n_b^1 \delta^1}{N_b^{(1)}}\right)} \phi_{l 
 \,,\, l -\delta} .  
\end{equation*}
\subsection{Massless Dirac Zero
  Modes\label{sec:zero}}\label{Diraczero} 

First, consider open strings on a stack of branes with gauge symmetry
$U(N_a+N_b)$. The fields $\Psi$ corresponding to the open string
states transform in the adjoint representation of the gauge
group. Massless fermions in four dimensions arise from
massless fermionic states in ten dimensions satisfying the Dirac
equation
\begin{equation}\label{Diraceq}
i\sum_{l=1}^3\left(\Gamma^{\overline l}D_{ l}-\Gamma^lD_l^\dag\right)\Psi(\vec z)=0,
\end{equation}
with $\Gamma^l$ and $\Gamma^{\overline l}$ being elements of the six
dimensional Clifford algebra and 
\begin{equation*}
D_{l}\psi_{\epsilon^1\epsilon^2\epsilon^3}(\vec z)=\overline\partial_{
  l}\psi_{\epsilon^1\epsilon^2\epsilon^3}(\vec z)+\left[A_{\overline
    l},\psi_{\epsilon^1\epsilon^2\epsilon^3}(\vec z)\right]. 
\end{equation*}
Here, $\psi_{\epsilon^1\epsilon^2\epsilon^3}$ are the eight components
of the Dirac fermion $\Psi$ and $\epsilon^l\in\{\pm\}$ denotes the
spin under the three Cartan generators of $SO(6)$ (i.e.\ the
components of $SO(6)$ fundamental weights).
For a given $SO(6)$ weight there are $\left( N_a + N_b\right)^2$
components forming the adjoint representation of the gauge group.
Eq.\ (\ref{Diraceq}) leads to three equations for each fundamental
weight,
\begin{equation}\label{Diracoperator}
D_l^\alpha\psi_{\vec \epsilon} =0\quad\text{with }\quad \alpha =
\left\{ \begin{array}{c l c} 1 & \text{if} & \epsilon^l = + \\
\dagger &\text{if} & \epsilon^l = - \end{array}\right\} \quad
\text{and}\quad l\in\{1,2,3\}.
\end{equation}
After turning on magnetic flux as in
(\ref{bifundamentalfieldstrength}), the states in $\psi_{\epsilon^l}$
decompose into the adjoint representation of $U(N_a)$ and $U(N_b)$ and
bifundamentals of $U(N_a)\times U(N_b)$ \cite{Cremades:2004wa}. 
The bifundamentals will be denoted by $\phi_{\vec \epsilon}$. For
e.g.\ $\epsilon_l = +$ the corresponding equation in
(\ref{Diracoperator}) reads
\begin{equation}
\overline\partial_{\overline l}\phi_{\vec \epsilon}+\frac{\pi \tilde
  I_{ab}^{(l)}}{2\mathrm{Im}(K_l)}z_l\phi_{\vec \epsilon}=0, 
\label{eq:dequ}
\end{equation}
As in \cite{Cremades:2004wa}, normalisable solutions to (\ref{eq:dequ})  for
fermions in  
the $\left({\mathbf N_{\mathbf a},\overline{\mathbf N}_{\mathbf b}}\right)$
will be considered. Normalisability leads to the condition
\begin{equation}
\epsilon_i = sign\left( I_{ab}^{(i)}\right) ,\,\,\, i \in \left\{
  1,2,3\right\} .
\label{eq:nobility}
\end{equation}
The chirality of the resulting four dimensional massless fermion is
fixed by the sign of $\epsilon^1\epsilon^2\epsilon^3$, i.e.\ by
$sign\left(\tilde{I}_{ab}^{(1)}\tilde{I}_{ab}^{(2)}\tilde{I}_{ab}^{(3)}\right)$. 
Apart from that the solutions depend only on the absolute values
$\left| I_{ab}^{(i)}\right|$. In the following vertical bars will be
dropped and postive $I_{ab}^{(i)}$'s will be assumed since negative
values can be accomodated easily by changing the chirality. The
following ansatz solves (\ref{eq:dequ})
\begin{equation}
\left(\phi_{\vec \epsilon}\left(\vec
z\right)\right)_{k_ak_b}=
\,\mathrm{e}^{\text{i}\pi\,\sum_{l=1}^3 \frac{\tilde 
    I_{ab}^{(l)}}{\mathrm{Im}(K_l)}\, z_l\mathrm{Im}(z_l)}\,
\xi_{k_ak_b}\left(\vec z\right)  . 
\label{eq:zmode}
\end{equation}
The
$\xi_{k_ak_b}$ are holomorphic functions of the $z^i$. Plugging this
ansatz into the boundary conditions (\ref{phicomptrafo}) yields
\begin{align}
\xi_{k_a,k_b}(z_1+1,z_2,z_3)&= \mathrm{e}^{2\pi \text{i}\left(\tilde
I_{ab}^{(1)}l^1+\frac{n_b^1\delta^1}{N_b^{(1)}}\right)}\xi_{k_a,k_b} 
\left( \vec z\right),\label{eq:bc1} \\
\xi_{k_a,k_b}(z_1+K_1,z_2-K_2,z_3)&= \mathrm{e}^{-\pi \text{i}\tilde
I_{ab}^{(1)}\left(2z_1+K_1\right)}\mathrm{e}^{\pi i\tilde
I_{ab}^{(2)}\left(2z_2-K_2\right)}\xi_{k_a+(1,-1,0),k_b+(1,-1,0)}
\left(\vec z\right), \label{eq:bc2} \\
\xi_{k_a,k_b}(z_1,z_2+1,z_3)&= \mathrm{e}^{2\pi \text{i}\left(\tilde 
I_{ab}^{(2)}l^2+
-\frac{m_b^2\delta^2}{N_b^{(2)}}\right)}
\xi_{k_a,k_b}(\vec z),\label{eq:bc3} \\
\xi_{k_a,k_b}(z_1,z_2+K_2,z_3-K_3)&= \mathrm{e}^{-\pi \text{i}\tilde
I_{ab}^{(2)}\left(2z_2+K_2\right)}\mathrm{e}^{\pi i\tilde
I_{ab}^{(3)}\left(2z_3-K_3\right)}\xi_{k_a+(0,1,-1),k_b+(0,1,-1)}\left(\vec
z\right),\label{eq:bc4} \\
\xi_{k_a,k_b}(z_1,z_2,z_3+1)&= \mathrm{e}^{2\pi \text{i}\left(\tilde I_{ab}^{(3)}
l^3-\frac{m_b^3\delta^3}{N_b{(3)}}\right)}
\xi_{k_a,k_b}(\vec z),\label{eq:bc5} \\
\xi_{k_a,k_b}(z_1,z_2,z_3+2K_3)&= \mathrm{e}^{-4\pi \text{i}\tilde 
I_{ab}^{(3)}\left(z_3+K_3\right)}
\xi_{k_a+(0,0,2),k_b+(0,0,2)}\left(\vec z\right) .
\label{eq:bc6}
\end{align}
First, focus on boundary conditions (\ref{eq:bc1}), (\ref{eq:bc3}),
(\ref{eq:bc5}), resulting in the general solution
\begin{equation}
\xi_{k_a ,k_b}\left(\vec z\right) =  
\sum_{\vec n \in {\mathbb Z}^3}\text{e}^{2\pi \text{i} \sum_{k=1}^3\left( n^k +
\tilde{I}_{ab}^{(k)}l^k+\phi^{(k)}\right)z^k} \rho_{\vec n}
\left( \vec l\right) ,
\label{eq:solzp1}
\end{equation}
where
\begin{align}
\phi^{(1)} & = -\frac{n_b^1\left(M_b Q\delta\right)^1
  +m_b^1\left(M_aP\delta\right)^1} 
{N_b^{(1)}},\,\\ \phi^{(i)}
& = \frac{m_b^i\left(M_bQ\delta\right)^i + n_b^i\left(M_aP\delta\right)^i}
{N_b^{(i)}}\,\,\, \text{for}\,\,\, i\in\left\{
  2,3\right\} .
\end{align}
On the right-hand sides of (\ref{eq:bc1}), (\ref{eq:bc3}),
(\ref{eq:bc5}) there will be additional, trivial phase factors of the
form $\text{exp}\left[ 2\pi\text{i}I\right]$ with the integer $I$ given by
$\left( M_a P\delta\right)^1$, $-\left( M_aP \delta\right)^2$, $-\left(
  M_aP\delta\right)^3$, respectively. The insertion of these factors
of one will be helpful in mapping zero mode labels to intersection
labels on the type IIA side, shortly.
The last term, $\rho_{\vec n}\left(\vec l\right)$ stands for $\vec z$
independent factors which will be 
further fixed by solving the remaining boundary
conditions. Imposing conditions (\ref{eq:bc2}), (\ref{eq:bc4}),
(\ref{eq:bc6}) and comparing coefficients at coinciding powers of $\exp
z^i$ leads to
\begin{align}
\frac{\rho_{\vec n}\left( \vec l + \left(1,-1,0\right)^T\right)
}{\rho_{\vec n}\left( \vec l\right) } & = \text{e}^{2\pi \text{i}\left\{
      \left( n^1 +\tilde{I}_{ab}^{(1)}\left( l^1+\frac{1}{2}\right)
        +\phi^{(1)}\right) K_1 -
\left( n^2 +\tilde{I}_{ab}^{(2)}\left( l^2-\frac{1}{2}\right)
        +\phi^{(2)}\right) K_2\right\}} ,\label{eq:rbc1}\\
\frac{\rho_{\vec n}\left( \vec l + \left(0,1,-1\right)^T\right)
}{\rho_{\vec n}\left( \vec l\right) } & = \text{e}^{2\pi \text{i}\left\{
      \left( n^2 +\tilde{I}_{ab}^{(2)}\left( l^2+\frac{1}{2}\right)
        +\phi^{(2)}\right) K_2 -
\left( n^3 +\tilde{I}_{ab}^{(3)}\left( l^3-\frac{1}{2}\right)
        +\phi^{(3)}\right) K_3\right\}} ,\label{eq:rbc2}\\
\frac{\rho_{\vec n}\left( \vec l + \left(0,0,2\right)^T\right)
}{\rho_{\vec n}\left( \vec l\right) } & = \text{e}^{4\pi \text{i}\left\{
      \left( n^3 +\tilde{I}_{ab}^{(3)}\left( l^3+1\right)
+\phi^{(3)}\right) K_3\right\} } .\label{eq:rbc3}
\end{align}
These conditions are solved by
\begin{equation}
\rho_{\vec n}\left( \vec l\right) = {\cal N}_{\vec n} \prod_{l=1}^3
\text{exp}\left[\frac{\text{i}\pi \left( n^h + \tilde{I}^{(h)} _{ab} l^h
    +\phi^{(h)} \right) ^2K_h}{\tilde{I}_{ab}^{(h)}} \right] .
\label{eq:rhosol}
\end{equation}
Independent normalisation constants $ {\cal N}_n$ indicate independent
zero modes. Imposing invariance under shifts of $\vec l$ by elements
of $\Gamma_a \cap \Gamma_b$ identifies some constants. For
$\lambda \in \Gamma_a \cap \Gamma_b$ this leads to 
\begin{equation}
{\cal N}_{\vec n} = {\cal N}_{\vec n^\prime}\,\,\, ,\,\,\, \text{for}
  \,\,\,
{n^\prime}^i = n^i +\tilde{I}_{ab}^{(i)}\lambda^i .
\label{eq:consteq}
\end{equation}
Possible lattices $\Gamma_a \cap \Gamma_b$ are listed in appendix
\ref{ap:proof}. For all cases one finds
\begin{equation}
\text{number of independent constants} = \frac{I_{ab}^{(1)}
  I_{ab}^{(2)}I_{ab}^{(3)}}{d^{(1)}d^{(2)}d^{(3)}} 
.
\label{eq:nic} 
\end{equation}
For the final counting of zero modes it is worthwhile noticing that
boundary conditions relate different pairs $k_a , k_b$ of identical
$k_a - k_b$ (see (\ref{eq:bc1})-(\ref{eq:bc6})). In
(\ref{eq:rbc1})-(\ref{eq:rbc3}) this reflected by relating different
$l$'s but not different $\delta$'s. In conclusion, the number of
independent zero modes is given by multiplying the number of independent
constants (\ref{eq:nic}) times the number of inequavalent
$\delta$'s. Going through all examples in appendix \ref{ap:proof} one
finds
\begin{equation*}
\text{number of independent zero modes} = \frac{ I_{ab}^{(1)}
  I_{ab}^{(2)}I_{ab}^{(3)}}{2} .
\end{equation*}
As expected, this equals the intersection number in the T-dual type IIA
configuration. It will be useful to detail the relation between
intersections and zero modes by identifying their labellings. 
In \cite{Forste:2014bfa} intersections in type IIA theory are labelled
by a triplet $j^{(1)}$, $j^{(2)}$, $j^{(3)}$ of the following form
\begin{equation}
j =\left( t_1 m_b^1 - t_2 n_b^2, t_3 m_b^2 - t_4 n_b^2, t_5 m_b^3 -
  t_6 n_b^3\right) ,\,\,\, \left( t_1,\ldots , t_6\right) \in 
\Lambda_{\text{SO(12)}}.
\label{eq:2al}
\end{equation}
This is subject to equivalence relations which will not be further
discussed since matching of the overall numbers has already been
established. 

The relation (\ref{eq:consteq}) is taken into account by renaming
the summation index
\begin{equation*}
n^i = \tilde{I}_{ab}\lambda^i  + k^i ,
\end{equation*}
where $k$ is a fixed label and $\lambda\in\Gamma_a\cap
\Gamma_b$ is summed over. Combining (\ref{eq:solzp1}) and
(\ref{eq:rhosol}) one obtains for one zero mode
\begin{equation*}
\xi^{k,\delta}_{l,l-\delta} = {\cal N}_k\sum_{\lambda \in
  \Gamma_a\cap\Gamma_b} \prod_{h=1}^3 \text{e}^{2 \pi
  \text{i}\tilde{I}_{ab}^h\left[ \left(\lambda^h + l^h + \frac{k^h +
    \phi^{(h)}}{\tilde{I}_{ab}^{(h)}}\right)z^h + \frac{K_h}{2}\left(
        \lambda^h + l^h+
\frac{ k^h +
    \phi^{(h)}}{\tilde{I}_{ab}^{(h)}}\right)^2\right]} ,
\end{equation*}
which is now labelled by the pair $k,\delta$. 
To make contact with the type IIA labelling one notices that the
solution depends only on the
combination $k+\phi$ which can be broought into the form
\begin{equation}
k + \phi = \frac{j}{N_b} .
\label{eq:2b2alab}
\end{equation}
where $j$ is the type IIA label (\ref{eq:2al}) with
\begin{align}
&t_1 = -\left( M_a P\delta\right)^1 +k^1\,\,\, ,\,\,\, t_2 = \left( M_b
  Q\delta\right)^1 - k^1\,\,\, ,\,\,\, t_3 = \left(
  M_bQ\delta\right)^2 + k^2 ,\nonumber \\
&t_4 = -\left( M_aP\delta\right)^2 - k^2\,\,\ ,\,\,\, t_5 =
  \left(M_bQ\delta\right)^3 + k^3 \,\,\, ,\,\,\, t_6 = -\left(
  M_aP\delta\right)^3 -k^3 .
\end{align}
That this is really in $\Lambda_{\text{SO(12)}}$ can be seen with
(\ref{eq:latemb}), (\ref{eq:DAB}) and the fact that $\delta \in
\Lambda_{\text{SO(6)}}$. Expressing the type IIB label in terms of the
type IIA label via (\ref{eq:2b2alab}) leads finally to
\begin{equation}
\xi^{j}_{l,l-\delta}\equiv \xi^{j}_l = {\cal N}_j^{ab}\sum_{\lambda \in
  \Gamma_a\cap\Gamma_b} \prod_{h=1}^3 \text{e}^{2 \pi
  \text{i}\tilde{I}_{ab}^h\left[ \left(\lambda^h + l^h + \frac{N_a^{(h)}j
    }{I_{ab}^{(h)}}\right) z^h + \frac{K_h}{2}\left(
        \lambda^h + l^h+
\frac{ N_a^{(h)}j
    }{I_{ab}^{(h)}}\right)^2\right]} ,
\label{eq:xisol}
\end{equation}    
where the notation has been changed to remove a redundancy in
specifying the $\delta$ dependence of the zero mode. The notation for
the original zero mode (\ref{eq:zmode})  will be changed accordingly
$\phi_{k_ak_b} \to \phi^i _l$. Keep in mind that in 
(\ref{eq:zmode}) and (\ref{eq:xisol}) one should actually replace, $I_{ab}^h \rightarrow \left| I_{ab}^h\right|$ (see discussion after (\ref{eq:nobility})).

\subsection{Normalisation Factor}

In order to get canonically normalised kinetic terms in four dimensions the zero modes need to satisfy the orthogonality relation \cite{Cremades:2004wa}\footnote{The factor ${\alpha^\prime}^{-3}$ has been included to match finally the convention of \cite{Cremades:2004wa}
 in which the gauge coupling is given by $\text{e}^{\Phi_b/2}{\alpha^\prime}^{3/2}$. 
 (At the moment $\alpha^\prime$ is fixed as in (\ref{eq:scalefix})).}
\begin{equation}\label{eq:ortho}
{\alpha^\prime}^{-3}\text{e}^{-\Phi_b}\prod_{l=1}^3 \frac{\text{Im}
    \tau_l}{\text{Im} K_l \left| 1 -\tau_l\right|^2}\int_{T^6}\mathrm{d}^6z\,
    \mathrm{Tr}\left\{\phi^i\cdot\left(\phi^j\right)^\dag
\right\}=\delta_{i,j}\,,
\end{equation}
where the integration is over complex coordinates (\ref{eq:compco}) and the metric is taken from (\ref{eq:metz}). Further, $\Phi_b$ is the ten dimensional type IIB dilaton which is 
chosen to be constant. Its exponential in (\ref{eq:ortho}) is a universal 
factor at all open string tree level contributions to the effective action. 
The domain of integration in (\ref{eq:ortho}) is given by the fundamental domain of $T^6$, which is the unit cell of the lattice spanned by
\begin{align}
      \vec v_1=(1,0,0)^T\,,\quad \vec v_2=(K_1,-K_2,0)^T\,,\quad \vec v_3=(0,1,0)^T\,,\\\nonumber
       \vec v_4=(0,K_2,-K_3)^T\,,\quad \vec v_5=(0,0,1)^T\,,\quad \vec v_6=(0,0,2K_3)^T\,.
\end{align}
With that the integration in (\ref{eq:ortho}) can be expressed as
\begin{equation*}
  \prod_{l=1}^3 \frac{\text{Im}
    \tau_l}{\text{Im} K_l \left| 1 -\tau_l\right|^2}  \int_{T^6}\mathrm{d}^6z=2\prod_{l=1}^3
    \frac{\mathrm{Im}(\tau_l)}{|1-\tau_l|^2}
    \int_0^1\mathrm{d}y_{2l-1}\int_0^1\mathrm{d}y_{2l}\,.
\end{equation*}
For each zero mode  $\phi^i$ and $\phi^j$ 
the parameters $\delta_{i}\,,\delta_{j}\in\frac{\Lambda_{\text{SO}(6)}}{\Gamma_d}$
are fixed according to the definition of the labels in (\ref{eq:2b2alab})
(see also the discussion after (\ref{eq:nic})). 
Hence, the sum in the trace in (\ref{eq:ortho}) has to be taken only over $l\in\frac{\Lambda_{\text{SO}(6)}}{\Gamma_a\cap\Gamma_b}$  
\begin{equation}
    \mathrm{Tr}\left\{\phi^i
    \cdot\left(\phi^j\right)^\dag\right\}=
    \sum_{l\in\frac{\Lambda_{\text{SO}(6)}}{\Gamma_a\cap\Gamma_b}}\,
    \delta_{\delta_i,\delta_j}\,\phi_{l,l-\delta_i}^i\cdot
    \left(\phi^j_{l,l-\delta_j}\right)^\dag\,,
\label{eq:tracesuml}
\end{equation}
where the Kronecker delta $\delta_{\delta_i,\delta_j}$ ensures that the sum is indeed a trace. The product of wavefunctions $\phi_{l,l-\delta_i}^i\cdot\left(\phi^j_{l,l-\delta_j}\right)^\dag$ satisfies the following boundary conditions
\begin{align}\nonumber
    \phi_{l,l-\delta_i}^i
    \left(\phi^j_{l,l-\delta_j}\right)^\dag(..., y_2+1,...)&=\phi_{l+(1,-1,0),l+(1,-1,0)
    -\delta_i}^i
    \left(\phi^j_{l+(1,-1,0),l+(1,-1,0)-\delta_j}\right)^\dag(..., y_2,...)\,,\\\nonumber
     \phi_{l,l-\delta_i}^i
     \left(\phi^j_{l,l-\delta_j}\right)^\dag(..., y_4+1,...)&=
     \phi_{l+(0,1,-1),l+(0,1,-1)-\delta_i}^i
     \left(\phi^j_{l+(0,1,-1),l+(0,1,-1)-\delta_j}\right)^\dag(..., y_4,...)
     \,,\\
      \phi_{l,l-\delta_i}^i
      \left(\phi^j_{l,l-\delta_j}\right)^\dag(..., y_6+1)&=\phi_{l+(0,0,2),l+(0,0,2)-\delta_i}^i
      \left(\phi^j_{l+(0,0,2),l+(0,0,2)-\delta_j}\right)^\dag(..., y_6)\,,
\end{align}
and therefore integrals of 
$\phi_{l,l-\delta_i}^i\cdot\left(\phi^j_{l,l-\delta_j}\right)^\dag$ 
over $T^6$, 
with different values for $l$ can be related to $T^6$ lattice shifts in the following way,
\begin{align}
    \int_0^1\mathrm{d}y_2\,\phi_{l+(1,-1,0),l+(1,-1,0)-\delta_i}^i
    \left(\phi^j_{l+(1,-1,0),l+(1,-1,0)-\delta_j}\right)^\dag&= \int_1^2\mathrm{d}y_2\,\phi_{l,l-\delta_i}^i
    \left(\phi^j_{l,l-\delta_j}\right)^\dag\,,
    \nonumber\\
       \int_0^1\mathrm{d}y_4\,\phi_{l+(0,1,-1),l+(0,1,-1)-\delta_i}^i
       \left(\phi^j_{l+(0,1,-1),l+(0,1,-1)-\delta_j}\right)^\dag&= \int_1^2\mathrm{d}y_4\,\phi_{l,l-\delta_i}^i
       \left(\phi^j_{l,l-\delta_j}\right)^\dag\,,  \label{eq:shiftfundom}\\
        \int_0^1\mathrm{d}y_6\left(\phi^j_{l+(0,0,2),l+(0,0,2)-\delta_j}\right)^\dag&= \int_1^2\mathrm{d}y_2\,\phi_{l,l-\delta_i}^i
        \left(\phi^j_{l,l-\delta_j}\right)^\dag.  
        \nonumber
\end{align}
The relations (\ref{eq:shiftfundom}) can be used to replace the sum over $l$ by an enlarged domain of integration. That means instead of integrating all terms, 
belonging to the trace, over the fundamental domain of $T^6$, we just need to 
integrate one term with a fixed $l$, for example $l=0$, over the enlarged domain 
of integration $\tilde C$, where $\tilde C$ is given by the unit cell of the 
lattice spanned by $\vec v_1=(1,0,0)^T$, $\vec v_3=(0,1,0)^T$ 
and $\vec v_5=(0,0,1)^T$ as before, but 
\begin{align}\nonumber
    \vec v_2=\begin{pmatrix}\frac{N_a^{(1)}N_b^{(1)}}{d^{(1)}}K_1\\0\\0\end{pmatrix}\,,\, \vec v_4=\begin{pmatrix}0\\\frac{N_a^{(2)}N_b^{(2)}}{d^{(2)}}K_2\\0\end{pmatrix}\,,\, \vec v_6=\begin{pmatrix}0\\0\\\frac{N_a^{(3)}N_b^{(3)}}{d^{(3)}}K_3\end{pmatrix},\,\text{for $\Gamma_a\cap\Gamma_b=\Gamma_1$}\,,\\
     \vec v_2=\begin{pmatrix}\frac{N_a^{(1)}N_b^{(1)}}{2d^{(1)}}K_1\\\frac{N_a^{(2)}N_b^{(2)}}{d^{(2)}}K_2\\0\end{pmatrix}\,,\, \vec v_4=\begin{pmatrix}\frac{N_a^{(1)}N_b^{(1)}}{2d^{(1)}}K_1\\-\frac{N_a^{(2)}N_b^{(2)}}{d^{(2)}}K_2\\0\end{pmatrix}\,,\, \vec v_6=\begin{pmatrix}0\\0\\\frac{N_a^{(3)}N_b^{(3)}}{d^{(3)}}K_3\end{pmatrix},\,\text{for $\Gamma_a\cap\Gamma_b=\Gamma_2$}\,,\\
        \vec v_2=\begin{pmatrix}\frac{N_a^{(1)}N_b^{(1)}}{2d^{(1)}}K_1\\\frac{N_a^{(2)}N_b^{(2)}}{d^{(2)}}K_2\\0\end{pmatrix}\,,\, \vec v_4=\begin{pmatrix}\frac{N_a^{(1)}N_b^{(1)}}{2d^{(1)}}K_1\\-\frac{N_a^{(2)}N_b^{(2)}}{d^{(2)}}K_2\\0\end{pmatrix}\,,\, \vec v_6=\begin{pmatrix}0\\\frac{N_a^{(2)}N_b^{(2)}}{d^{(2)}}K_2\\\frac{N_a^{(3)}N_b^{(3)}}{d^{(3)}}K_3\end{pmatrix},\,\text{for $\Gamma_a\cap\Gamma_b=\Gamma_3$},\nonumber
\end{align}
where e.g.\ $\Gamma_a\cap\Gamma_b = \Gamma_1$ means that it is of the form $\Gamma_1$ in Appendix
\ref{ap:proof}.

The explicit expression for
$\phi_{0,-\delta_i}^i\cdot\left(\phi_{0,-\delta_j}^j\right)^\dag$ 
can be deduced by inserting (\ref{eq:zmode}) and (\ref{eq:xisol}),
\begin{align}\nonumber
    \phi_{0,-\delta_i}^i\cdot\left(\phi_{0,-\delta_j}^j\right)^\dag
    =&\,\mathcal{N}^{ab}_i{\mathcal{N}^\star}^{ab}_j\mathrm{exp}\left\{-2\pi\sum_{k=1}^3\frac{\tilde I_{ab}^{(k)}}{\mathrm{Im}(K_k)}\left(\mathrm{Im}(z_k)\right)^2\right\}
    \\& \hspace*{-1.8cm}
    \sum_{\lambda\in\Gamma_a\cap\Gamma_b}\sum_{\rho\in\Gamma_a\cap\Gamma_b}\prod_{h=1}^3
 \mathrm{exp}\left\{2\pi i \left[\left(\tilde I_{ab}^{(h)}\lambda_{ab}^{(h)}+\frac{i^{(h)}}{N_b^{(h)}}\right)z_h-\left(\tilde I_{ab}^{(h)}\rho_{ab}^{(h)}+\frac{j^{(h)}}{N_b^{(h)}}\right)\overline z_h\right]\right\}
 \nonumber\\
    & \hspace*{-1.8cm}
    \cdot\mathrm{exp}\left\{\pi i \left[\left(\tilde I_{ab}^{(h)}\lambda_{ab}^{(h)}+\frac{i^{(h)}}{N_b^{(h)}}\right)^2\frac{K_h}{\tilde I_{ab}^{(h)}}-\left(\tilde I_{ab}^{(h)}\rho_{ab}^{(h)}+\frac{j^{(h)}}{N_b^{(h)}}\right)^2\frac{\overline K_h}{\tilde I_{ab}^{(h)}}\right]\right\}\,.
\label{eq:phiphi}
\end{align}
The $y_1$, $y_3$, $y_5$ dependence of the integrand is contained in factors
$\text{exp}\left( 2\pi\text{i}y_{2h -1} M_h\right)$ with
\begin{equation}
 M_h= \tilde I_{ab}^{(h)}\lambda_{ab}^{(h)}+\frac{i^{(h)}}{N_b^{(h)}}-\tilde I_{ab}^{(h)}\rho_{ab}^{(h)}-\frac{j^{(h)}}{N_b^{(h)}} .
 \label{eq:intfact}
 \end{equation}
A closer look at (\ref{eq:intfact}) reveals that, taking the trace condition $\delta_{\delta_i,\delta_j}$ into account, the terms are actually integer, 
because the potentially non integer part, which is according to (\ref{eq:2b2alab}) given by $\frac{\delta_{i^{(h)}}}{N_b^{(h)}}-\frac{\delta_{j^{(h)}}}{N_b^{(h)}}$, vanishes for $\delta_i=\delta_j$. Hence, the integration of (\ref{eq:phiphi}) over $y_1$, $y_3$ and $y_5$
yields one if all $M_h$ in (\ref{eq:intfact}) vanish and zero otherwise. This implies a non vanishing result only for $\rho_{ab} =\lambda_{ab}$ and $i=j$, establishing orthogonality 
of the zero modes. The final result of the $y_1$, $y_3$, $y_5$ integration is
\begin{align}
\int dy_1 dy_3 dy_5 \, \phi_{0,-\delta_i}^i\cdot
\left(\phi_{0,-\delta_j}^j\right)^\dag = & \delta_{ij} \left|\mathcal{N}^{ab}_i\right|^2\nonumber \\
  &  \hspace*{-5.5cm}
 \sum_{\lambda\in\Gamma_a\cap\Gamma_b}
  \prod_{h=1}^3\mathrm{exp}\left\{-2\pi \frac{\tilde I_{ab}^{(h)}}{\mathrm{Im}(K_h)}\left(\mathrm{Im}(z_h)
  +\lambda^{(h)}\mathrm{Im}(K_h)+\frac{i^{(h)}}{N_a^{(h)}}
  \frac{\mathrm{Im}(K_h)}{I_{ab}^{(h)}}\right)^2\right\}\,.
\end{align}
Similar to (\ref{eq:shiftfundom}), the sum over $\lambda$ can be replaced by an enlarged domain of 
integration over $y_2$, $y_4$, $y_6$ 
\begin{equation*}
    \int_{\tilde C}\mathrm{d}y_2\mathrm{d}y_4\mathrm{d}y_6\sum_{\lambda\in\Gamma_a\cap\Gamma_b}
    \ldots\rightarrow \int_{\mathbb{R}^3}\mathrm{d}y_2\mathrm{d}y_4\mathrm{d}y_6\ldots
    =  \frac{1}{2}  \int_{\mathbb{R}^3}\mathrm{d}\left(\frac{\mathrm{Im}(z_h)}{\mathrm{Im}(K_h)}\right)\ldots
\end{equation*}
There are three remaining Gaussian integrals, solved by
\begin{align}
  \prod_{h=1}^3\int_{\mathbb{R}^3}\mathrm{d}\left(\mathrm{Im}(z_h)\right)\,\mathrm{e}^{-2\pi \frac{\tilde I_{ab}^{(h)}}{\mathrm{Im}(K_h)}\left(\mathrm{Im}(z_h)+\frac{i^{(h)}}{N_a^{(h)}}\frac{\mathrm{Im}(K_h)}{I_{ab}^{(h)}}\right)^2}=\prod_{h=1}^3\sqrt{\frac{\mathrm{Im}(K_h)}{2\tilde I_{ab}^{(h)}}}\,.
\end{align}
Plugging the results into (\ref{eq:ortho}), one finds the normalisation condition 
\begin{equation}
 {\alpha^\prime}^{-3}\text{e}^{-\Phi_b}  \left|\mathcal{N}_i^{ab}\right|^2
 \prod_{h=1}^{3}\frac{\mathrm{Im}(\tau_h)}{|1-\tau_h|^2}\left(2\tilde I_{ab}^{(h)}\mathrm{Im}(K_h)\right)^{-\frac{1}{2}}= 1\,.
\label{eq:normcond}    
\end{equation}
In the next section, normalisation factors will be real solutions of (\ref{eq:normcond}).

\section{Yukawa Couplings}

The configuration considered in this section will be the T-dual of
type IIA with three stacks of intersecting D6 branes. That is, 
(\ref{bifundamentalfieldstrength}) will be amended to\footnote{Now
  also $N_c =N_c^{(1)}N_c^{(2)}N_c^{(3)}/2$.} 
\begin{align}
& F_{z_1\overline z_1}=\frac{\pi
  i}{\mathrm{Im}\left(K_1\right)} \begin{pmatrix}
  \frac{n^1_a}{N^{(1)}_a}\mathbbm{1}_{N_a}   
  \\
& \frac{n^1_b}{N^{(1)}_b}\mathbbm{1}_{N_b}\\
& & \frac{n^1_c}{N^{(1)}_c}\mathbbm{1}_{N_c}\end{pmatrix},\nonumber \\
& F_{z_2\overline z_2}=-\frac{\pi
  i}{\mathrm{Im}\left(
  K_2\right)}\begin{pmatrix}\frac{m^2_a}{N^{(2)}_a}\mathbbm{1}_{N_a}&  
  \\ &\frac{m^2_b}{N^{(2)}_b}\mathbbm{1}_{N_b}\\
& &\frac{m^2_c}{N^{(2)}_c}\mathbbm{1}_{N_c}\end{pmatrix},\\\nonumber
& F_{z_3\overline z_3}=-\frac{\pi
  i}{\mathrm{Im}\left
  (K_3\right)}\begin{pmatrix}\frac{m^3_a}{N^{(3)}_a}\mathbbm{1}_{N_a} 
  &\\&\frac{m^3_b}{N^{(3)}_b}\mathbbm{1}_{N_b}\\
& &\frac{m^3_c}{N^{(3)}_c}\mathbbm{1}_{N_c}\end{pmatrix}.
\end{align}
This breaks the original $U\left( N_a N_b N_c\right)$ gauge symmetry
to $U\left( N_a\right)\times U\left( N_b\right) \times U\left(
  N_c\right)$ which is further broken by Wilson lines to $U(1)^3$. 

\subsection{Two Extra Dimensions}

It will be useful to recapitulate and to generalise the computation of
Yukawa couplings in the case of two extra dimensions. This has been
dealt with in \cite{Cremades:2004wa} for the case that all pairs from
$\left\{ N_a, N_b , N_c\right\}$ are coprime. The computation of the
Yukawa coupling boils down to evaluating integrals of the form\footnote{For simplicity, 
moduli dependence will be suppressed in the present discussion.}
\begin{equation}
\left|\lambda_{ijk}\right| = \int_{T^2} d^2z\sum_{k_a=0}^{N_a
  -1}\sum_{k_b=0}^{N_b-1}\sum_{k_c=0}^{N_c-1}
\phi^{i,I_{ab}}_{k_ak_b}\phi^{j,I_{ca}}_{k_ck_a}{\phi^\star}^{k,I_{cb}}_{k_bk_c} .  
\label{eq:t2cop}
\end{equation}
The $\phi$'s denote zero modes in bifundamentals as before. Now, the
zero mode label has been supplemented by the intersection number. For
all $N_\alpha$'s being coprime the matrix elements are related by shifts
by cycles of the $T^2$ (analogous to e.g.\ expression
(\ref{eq:bc2})). This enabled the authors of \cite{Cremades:2004wa} to
trade the sums 
(\ref{eq:t2cop}) for an enlarged integration region
$\tilde{T}^2$. Before outlining more details it will be uesful to
include also the discussion of non coprime pairs among the $N_\alpha$'s. 
In this case there are subsets within all matrix elements
invariant under shifting zero mode arguments by $T^2$ cycles. As
discussed in (\ref{eq:kdiff}) these 
sectors are characterised by differences in row and column number, e.g.\
\begin{equation}
k_a - k_b = \delta_{ab}\,\,\,\text{mod} \,\,\, d_{ab}
\label{eq:diffcla}
\end{equation}
where
\begin{equation}
d_{\alpha\beta} = g.c.d.\left( N_\alpha , N_\beta\right) ,\,\,\,
\text{for}\,\,\, \alpha, \beta \in \left\{ a,b,c\right\} .
\label{eq:gcdab}
\end{equation}
Different $\delta_{\alpha\beta}$'s belong to different zero modes. Expression
(\ref{eq:t2cop}) should be modified to
\begin{equation}
\left|\lambda_{ijk}\right| = \int_{T^2} d^2z\sum_{k_a=0}^{N_a
  -1}\sum_{k_b=0}^{N_b-1}\sum_{k_c=0}^{N_c-1}\delta_{k_a - k_b,
  \delta_{ab}} \delta_{k_c- k_a ,\delta_{ca}}\delta_{k_b- k_c,
  \delta_{bc}}
\delta_{\delta_{ab}+\delta_{bc}+\delta_{ca}}  
\phi^{i,I_{ab}}_{k_ak_b}
\phi^{j,I_{ca}}_{k_ck_a}
{\phi^\star}^{k,I_{cb}}_{k_bk_c}
.
\label{eq:mod2d}   
\end{equation}
Here, the first three $\delta$'s are usual Kronecker deltas on
${\mathbb Z}_{d_{ab}}$, e.g.\ the first is one if (\ref{eq:diffcla})
holds and zero otherwise. The last $\delta$ ensures that the trace is
taken and is defined as 
\begin{equation}
\delta_{\rho}=\left\{ \begin{array}{l l}
1 & \,\,\,\text{for}\,\,\, \rho = 0 \,\,\, \text{mod} \,\,\,
g.c.d.\left( d_{ab}, 
    d_{bc}, d_{ca}\right) , \\
0 & \,\,\,\text{else.} \end{array}\right.
\label{eq:diracdef}
\end{equation}
The following
abreviations will be convenient. Similar to the greatest common
divisor (\ref{eq:gcdab}) the lowest common multiple will be denoted as
\begin{equation*}
N_{\alpha\beta} = l.c.m.\left( N_\alpha, N_\beta\right)\,\,\,
\text{for}\,\,\, \alpha , \beta \in \left\{ a,b,c\right\} .
\end{equation*}
For $d_{abc}$ given by
\begin{equation}
d_{abc} = g.c.d.\left( N_{ab}, N_c\right) = g.c.d.\left( N_{ca},
  N_b\right) = g.c.d.\left( N_{bc}, N_a\right) 
\label{eq:dabc}
\end{equation}
one finds
\begin{equation*}
d_{abc} = \left\{ \begin{array}{l l}
d_{ab} d_{bc} d_{ca} &\,\,\,\text{for}\,\,\, d_{ab}\not= d_{bc},\,\,\,
                       d_{ab}\not= d_{ca},\,\,\, d_{bc} \not= d_{ca}
                       ,\\
d_{ab}d_{bc} & \,\,\,\text{for}\,\,\, d_{ab}=d_{ac}\not= d_{bc} ,\\
d_{ab}^2 & \,\,\,\text{for}\,\,\, d_{ab}=d_{bc}=d_{ca} , \end{array} 
\right.
\end{equation*}
where cases which can be obtained by  permuations of $\left(
  a,b,c\right)$ have not been explicitly written. With $d_{abc}$ one
can relate the product of three numbers to its lowest common multiple
\begin{equation*}
N_a N_b N_c = d_{abc}\, l.c.m.\left( N_a, N_b, N_c\right) .
\end{equation*}
The double index e.g.\ $k_a, k_b$ can now be replaced by a single
index $l$ as in (\ref{eq:phasfac}) where it proves useful to change
notation slightly. For a fixed $\delta_{\alpha\beta}$ which is encoded
in the label $i$ one replaces
\begin{equation*}
\phi^{i,I_{\alpha\beta}}_{k_a k_b} =\phi^{i,I_{\alpha\beta}}_l \,\,\, , \,\,\, l
\in {\mathbb Z}_{N_{\alpha\beta}}.
\end{equation*}
With that notation, the Yukawa coupling (\ref{eq:mod2d}) reads
\begin{equation}
\left|\lambda_{ijk}\right| = \int_{T^2}d^2z\sum_{l=0}^{N_{abc}-1} \phi^{i,I_{ab}}_{l}
\phi^{j,I_{ca}}_{l-\delta_{ab}}
{\phi^\star}^{k,I_{cb}}_{l+\delta_{bc}}    .
\label{eq:moddt2}
\end{equation}
Notice, that it has been possible to drop the first three Kronecker
deltas of (\ref{eq:mod2d}). However, the last $\delta$ function in
(\ref{eq:mod2d}) translates into a selection rule involving the labels
$i,j,k$. Its explicit form depends on the so far unspecified way
$\delta_{\alpha\beta}$ is encoded in the label. Therefore it has been
left out in (\ref{eq:moddt2}) but should be kept in mind. 
The gauge indices (summation labels) have been chosen such that the
sum implies matrix multiplication, i.e.\ consecutive row and column
indices match (see (\ref{eq:soofnice})).  
Analogous to e.g.\ (\ref{eq:bc2}) $l$ can be shifted by one when
replacing $z \to z + \tau$, where $\tau$ is the complex structure
modulus of the compactification $T^2$. Any factor induced by such
shifts (cf (\ref{eq:biembed})) drops out due to the identity
\begin{equation}
\tilde{I}_{ab} +\tilde{I}_{bc} +\tilde{I}_{ca} = 0 . 
\label{eq:relationIab}
\end{equation} 
Therefore, one can replace the sum over $l$ by an enlarged integration
region leading to
\begin{equation}
\left|\lambda_{ijk}\right| = \int_{\tilde{T}^2}d^2z \phi^{i,I_{ab}}_{0}
\phi^{j,I_{ca}}_{-\delta_{ab}}
{\phi^\star}^{k,I_{cb}}_{\delta_{bc}}    ,
\label{eq:fint2}
\end{equation}
where $\tilde{T}_2$ has complex structure $N_{abc}\tau$.
From hereon one can use the techniques presented in
\cite{Cremades:2004wa} to complete the computation for the generalised
configuration with two extra dimensions.  

\subsection{Yukawa Couplings for the T-dual of
  $\mathbf{T^6_{\text{SO}(12)}}$}  

Now the Yukawa coupling is determined via computing\footnote{The sign is determined  exactly as in the factorisable case \cite{Cremades:2004wa} and not discussed here.}
\begin{equation*}
\left|\lambda_{ijk}\right| ={\alpha^\prime}^{-3}\text{e}^{-\Phi_b}\prod_{l=1}^3 \frac{\text{Im}
    \tau_l}{\text{Im} K_l \left| 1 -\tau_l\right|^2}\int_C d^6
z\sum_{l\in\frac{\Lambda_{\text{SO(6)}}}{\Gamma_a \cap \Gamma_b\cap
      \Gamma_c}} \phi^{j,I_{ca}}_{l} 
\phi^{i,I_{ab}}_{l-\delta_{ca}}
{\phi^\star}^{k,I_{cb}}_{l+\delta_{cb}}    ,
\end{equation*}  
where $z$ has been introduced in (\ref{eq:compco}), the prefactor
comes from $\sqrt{G}$ with the metric taken from (\ref{eq:metz}). The
region of integration is a  parallelepiped $C\subset {\mathbb C}^3$ whose
edges are given by the following vectors
\begin{equation}
\begin{array}{lll}
l_1^T = \left( 1,0,0\right) , & l_3^T =\left( 0,1,0\right) & l_5^T
                                                               =\left(
                                                               0,0,1\right)
                                                               ,\\
l_2 ^T =\left( K_1, -K_2 ,0\right) , & l_4^T =\left( 0, K_2 ,
                                         -K_3\right) , & l_6^T=\left(
                                                         0,0,2K_3\right) .
\end{array}
\label{eq:sparalped}
\end{equation}
Again, the sum over gauge indicies $l$ can be replaced by an enlarged
integration region since shifts by $l_2$, $l_4$ or $l_6$ induce index
shifts according to (\ref{eq:bc2}), (\ref{eq:bc4}), (\ref{eq:bc6}). To
be more specific, one needs to identify 
$\Gamma_a \cap \Gamma_b\cap \Gamma_c$. Repeating the analysis given
in appendix \ref{ap:proof} one finds that $\Gamma_a \cap \Gamma_b\cap
\Gamma_c$ is given by either $\Gamma_1$, $\Gamma_2$ or $\Gamma_3$ with 
$$
N_x^{(l)} = N_{abc}^{(l)}\equiv \frac{N_a^{(l)}N_b^{(l)}
  N_c^{(l)}}{d^{(l)}_{abc}} . 
$$
Here, $d_{\alpha\beta}^{(l)}$, $d_{abc}^{(l)}$ are defined as in
respectively (\ref{eq:gcdab}), (\ref{eq:dabc}) for each $l\in
\left\{ 1,2,3\right\}$. 
The summation over $l\in \frac{\Lambda_{\text{SO(6)}}}{\Gamma_a \cap \Gamma_b\cap
      \Gamma_c}$ can be traded for an integration over a larger parallelepiped 
    $\tilde{C}\in {\mathbb C}^3$,
\begin{equation*}
\left|\lambda_{ijk}\right| ={\alpha^\prime}^{-3}\text{e}^{-\Phi_b}\prod_{l=1}^3 \frac{\text{Im}
    \tau_l}{\text{Im} K_l \left| 1 -\tau_l\right|^2}\int_{\tilde{C}} d^6
z\,  \phi^{j,I_{ca}}_{0} 
\phi^{i,I_{ab}}_{-\delta_{ca}}
{\phi^\star}^{k,I_{cb}}_{\delta_{cb}}   .
\end{equation*}
The edges of $\tilde{C}$ are 
$l_1$, $l_3$ and $l_5$ as in (\ref{eq:sparalped}) but $l_2$, $l_4$ and
$l_6$ replaced 
by the generators of one of the lattices $\Gamma_1$, $\Gamma_2$ or
$\Gamma_3$ from appendix \ref{ap:proof} with
\begin{equation*}
N_x^{(l)} = N_{abc}^{(l)} K_l .
\end{equation*}
Next, the integration variables are replaced by $\left\{ y_1,\ldots
  ,y_6\right\}$ as in (\ref{eq:compco}),
\begin{equation*}
\left|\lambda_{ijk}\right| =2{\alpha^\prime}^{-3}\text{e}^{-\Phi_b}\prod_{l=1}^3 \frac{\text{Im}
    \tau_l}{\left| 1 -\tau_l\right|^2}\int_{\tilde{C}} d^6
y\, \phi^{j,I_{ca}}_{0} 
\phi^{i,I_{ab}}_{-\delta_{ca}}\
{\phi^\star}^{k,I_{cb}}_{\delta_{cb}}    .
\end{equation*}
The range for the $\left\{ y_1,y_3,y_5\right\}$ integration is the
cube spanned by $l_1$, $l_3$ and $l_5$. The range for $\left\{
y_2,y_4,y_6\right\}$ is a parallelepiped whose form depends on whether
$\Gamma_a\cap\Gamma_b\cap\Gamma_c$ is of the form $\Gamma_1$,
$\Gamma_2$, or $\Gamma_3$. One finds for the edges of the
parallelepiped 
\begin{align}
& l_2 =
  \frac{N_{abc}^{(1)}}{2}
\left(\begin{array}{c}2\\2\\1\end{array}\right),\,\,\,  
l_4 = \frac{N_{abc}^{(2)}}{2}
\left( \begin{array}{c}0\\2\\1\end{array}\right) ,\,\,\,
l_6 = \frac{N_{abc}^{(3)}}{2}
\left( \begin{array}{c}0\\0\\1\end{array}\right)\nonumber\\  
&\text{if form of $\Gamma_a \cap \Gamma_b\cap \Gamma_c$ is $\Gamma_1$}, 
\nonumber \\
&l_2 =
  \frac{N_{abc}^{(1)}}{4}
\left(\begin{array}{c}2\\2\\1\end{array}\right)
+ \frac{N_{abc}^{(2)}}{2}
\left( \begin{array}{c}0\\2\\1\end{array}\right)
,\,\,\,
l_4 = \frac{N_{abc}^{(1)}}{4}
\left(\begin{array}{c}2\\2\\1\end{array}\right)
- \frac{N_{abc}^{(2)}}{2}
\left( \begin{array}{c}0\\2\\1\end{array}\right),\nonumber \\
&l_6 = \frac{N_{abc}^{(3)}}{2}
\left( \begin{array}{c}0\\0\\1\end{array}\right)\,\,\,
\text{if form of $\Gamma_a \cap \Gamma_b\cap \Gamma_c$ is $\Gamma_2$}, 
\label{eq:tildeCdomain}\\
&l_2 =
  \frac{N_{abc}^{(1)}}{4}
\left(\begin{array}{c}2\\2\\1\end{array}\right)
+ \frac{N_{abc}^{(2)}}{2}
\left( \begin{array}{c}0\\2\\1\end{array}\right)
,\,\,\,
l_4 = \frac{N_{abc}^{(1)}}{4}
\left(\begin{array}{c}2\\2\\1\end{array}\right)
- \frac{N_{abc}^{(2)}}{2}
\left( \begin{array}{c}0\\2\\1\end{array}\right),\nonumber\\
&l_6 = \frac{N_{abc}^{(2)}}{2}
\left( \begin{array}{c}0\\2\\1\end{array}\right)
+\frac{N_{abc}^{(3)}}{2}
\left( \begin{array}{c}0\\0\\1\end{array}\right) \,\,\, 
\text{if form of $\Gamma_a \cap \Gamma_b\cap \Gamma_c$ is $\Gamma_3$}. 
\nonumber
\end{align}
Inserting the expressions from (\ref{eq:zmode}) and (\ref{eq:xisol}) into $\phi_0^{j,I_{ca}}\phi_{-\delta_{ca}}^{i,I_{ab}}\phi_{\delta_{cb}}^{k,I_{cb}}$, and using the relation (\ref{eq:relationIab}), with the choice
\begin{equation*}
 |\tilde I_{ab}^{(h)}|+|\tilde I_{ca}^{(h)}|=|\tilde I_{cb}^{(h)}|\,,
 \quad\text{and}\quad \tilde I_{ab}^{(h)}, I_{ca}^{(h)}, I_{cb}^{(h)}>0\,,
\end{equation*}
the explicit expression for the product of wavefunctions is given by
\begin{align}
\phi^{j,I_{ca}}_{0} 
\phi^{i,I_{ab}}_{-\delta_{ca}}\
{\phi^\star}^{k,I_{cb}}_{\delta_{cb}}
 = & 
 \,\,\,
  \mathcal{N}^{ca}_j\mathcal{N}^{ab}_i\mathcal{N}^{cb}_k
 \mathrm{e}^{-2\pi\sum_{h=1}^3\frac{\tilde I_{cb}^{(h)}}{\mathrm{Im}(K_h)}\left(\mathrm{Im}(z_h)\right)^2} \nonumber \\ &
 \hspace*{-3cm}
 \sum_{\lambda_{xy}\in\Gamma_x\cap\Gamma_y}
\prod_{h=1}^3
 \mathrm{e}^{2\pi i\left\{\left(\tilde I_{ab}^{(h)}\left(\lambda_{ab}^{(h)}
 -\delta_{ca}^{(h)}\right)+\tilde I_{ca}^{(h)}\lambda_{ca}^{(h)}+
 \frac{i^{(h)}}{N_b^{(h)}}+\frac{j^{(h)}}{N_a^{(h)}}\right)z_h-\left(\tilde I_{cb}^{(h)}\left(\lambda_{cb}^{(h)}-\delta_{ca}^{(h)}
 -\delta_{ab}^{(h)}\right)+\frac{k^{(h)}}{N_c^{(h)}}\right)\overline z_h\right\}}\\\nonumber
 & \hspace*{-3cm}\mathrm{e}^{\pi i\left\{\left(\tilde I_{ab}^{(h)}\left(\lambda_{ab}^{(h)}
 -\delta_{ca}^{(h)}\right)+\frac{i^{(h)}}{N_b^{(h)}}\right)^2\frac{K_h}{\tilde I_{ab}^{(h)}}+\left(\tilde I_{ca}^{(h)}\lambda_{ca}^{(h)}+\frac{j^{(h)}}{N_a^{(h)}}\right)^2\frac{\tau_h}{\tilde I_{ca}^{(h)}}-\left(\tilde I_{cb}^{(h)}\left(\lambda_{cb}^{(h)}
 -\delta_{ca}^{(h)}-\delta_{ab}^{(h)}\right)+\frac{k^{(h)}}{N_c^{(h)}}\right)^2\frac{\overline K_h}{\tilde I_{ca}^{(h)}}\right\}}\,,
\end{align}
where $x,y\in \left\{ a,b,c\right\}$
and the wavefunction $\phi_{k_b,k_c}^{k,I_{cb}}$ has been relabelled such that $k^{(h)}/N_b^{(h)}$ 
is replaced by $k^{(h)}/N_c^{(h)}$. (This corresponds to swapping the label of $\xi_{k_c,k_b}$ 
with minus the label of $\xi_{k_b,k_c}^\star$, see (\ref{eq:xisol}).)

Before performing the integration, a closer look at the terms
\begin{equation}
 \tilde I_{ab}^{(h)}\left(\lambda_{ab}^{(h)}-\delta_{ca}^{(h)}\right)+\tilde I_{ca}^{(h)}\lambda_{ca}^{(h)}+\frac{i^{(h)}}{N_b^{(h)}}+\frac{j^{(h)}}{N_a^{(h)}}-\tilde I_{cb}^{(h)}\left(\lambda_{cb}^{(h)}-\delta_{ca}^{(h)}-\delta_{ab}^{(h)}\right)-\frac{k^{(h)}}{N_c^{(h)}}\,,
\label{eq:termsw}
\end{equation}
reveals them to be integers.
From the way the labels $i,j,k$ in (\ref{eq:2b2alab}) where introduced, it can be deduced that 
the potentially non integer part in (\ref{eq:termsw}) is 
\begin{equation}
-  \tilde I_{ab}^{(h)}\delta_{ca}^{(h)}+\frac{\delta_{ab}^{(h)}}{N_b^{(h)}}
+\frac{\delta_{ca}^{(h)}}{N_a^{(h)}}+\tilde I_{cb}^{(h)}\left(\delta_{ca}^{(h)}+\delta_{ab}^{(h)}\right)
-\frac{\delta_{cb}^{(h)}}{N_c^{(h)}}\,.
\label{eq:termswnoninter}
\end{equation}
However, when considering the trace of
$\phi^{j,I_{ca}}_{l}\phi^{i,I_{ab}}_{l-\delta_{ca}}\phi^{k,I_{cb}}_{l+\delta_{cb}}$, only terms with $\delta_{ca}+\delta_{ab}=\delta_{cb}$ contribute and hence the terms 
in (\ref{eq:termswnoninter}) vanish, the expression in (\ref{eq:termsw}) is indeed integer $\forall\,\{\lambda_{ab},\,\lambda_{ca},\,\lambda_{cb}\}$ and $\forall\,h\in\{1,2,3\}$. Therefore 
the integration over $y_1$, $y_3$ and $y_5$ leads to Kronecker deltas,
\begin{align}
\int_0^1\mathrm{d}y_{2h-1}\,\mathrm{e}^{2\pi i \left( \tilde I_{ab}^{(h)}\left(\lambda_{ab}^{(h)}-\delta_{ca}^{(h)}\right)+\tilde I_{ca}^{(h)}\lambda_{ca}^{(h)}+\frac{i^{(h)}}{N_b^{(h)}}+\frac{j^{(h)}}{N_a^{(h)}}-\tilde I_{cb}^{(h)}\left(\lambda_{cb}^{(h)}-\delta_{ca}^{(h)}-\delta_{ab}^{(h)}\right)-\frac{k^{(h)}}{N_c^{(h)}}\right)y_{2h-1}} = &\\\nonumber
 &\hspace*{-27em}\delta_{ \tilde I_{ab}^{(h)}\left(\lambda_{ab}^{(h)}-\delta_{ca}^{(h)}\right)+\tilde I_{ca}^{(h)}\lambda_{ca}^{(h)}+\frac{i^{(h)}}{N_b^{(h)}}+\frac{j^{(h)}}{N_a^{(h)}}-\tilde I_{cb}^{(h)}\left(\lambda_{cb}^{(h)}-\delta_{ca}^{(h)}-\delta_{ab}^{(h)}\right)-\frac{k^{(h)}}{N_c^{(h)}}}\,,
\end{align}
which imply the following Diophantine equations $h\in\left\{1,2,3\right\}$
\begin{equation}
  \tilde I_{ab}^{(h)}\left(\lambda_{ab}^{(h)}-\delta_{ca}^{(h)}\right)+\tilde I_{ca}^{(h)}\lambda_{ca}^{(h)}+\frac{i^{(h)}}{N_b^{(h)}}+\frac{j^{(h)}}{N_a^{(h)}}-\tilde I_{cb}^{(h)}\left(\lambda_{cb}^{(h)}-\delta_{ca}^{(h)}-\delta_{ab}^{(h)}\right)-\frac{k^{(h)}}{N_c^{(h)}}=0\,.
\label{eq:dio}
\end{equation}
In \cite{Forste:2014bfa} Diophantine equations arose from the requirement that projections of the intersecting D6 branes form closed triangles in each plane. These equations could be solved 
in terms of wrapping numbers after performing a relabelling of the intersection points,
\begin{equation}
 i^{(h)}\rightarrow\frac{i^{(h)}}{d_b^{(h)}}I_{cb}^{(h)}\,,\quad j^{(h)}\rightarrow\frac{j^{(h)}}{d_a^{(h)}}I_{ba}^{(h)}\,,\quad k^{(h)}\rightarrow\frac{k^{(h)}}{d_c^{(h)}}I_{ac}^{(h)}\,,
\label{eq:rel}
\end{equation}
where e.g.\ $d_a^{(h)} = g.c.d\left( I^{(h)}_{ab},I^{(h)}_{ac}\right)$. It can happen that intersection points lose
their label. The corresponding Yukawa couplings are equal to others for which no label is 
lost \cite{Forste:2014bfa}. Performing the same relabelling (\ref{eq:rel}) on the type IIB side,
the solutions to (\ref{eq:dio}) are given by
\begin{align}
 \lambda_{ab}^{(h)}&=N_{abc}^{(h)}p^{(h)}+N_{a}^{(h)}N_b^{(h)}M_c^{(h)}q^{(h)}
 +\frac{j^{(h)}}{d_a^{(h)}}N_b^{(h)}+\delta_{ca}^{(h)}\,,\nonumber\\
  \lambda_{ca}^{(h)}&=N_{abc}^{(h)}p^{(h)}+N_{a}^{(h)}M_b^{(h)}N_c^{(h)}q^{(h)}
  -\frac{k^{(h)}}{d_c^{(h)}}N_a^{(h)}\,,\label{eq:soldio}\\\nonumber
   \lambda_{cb}^{(h)}&=N_{abc}^{(h)}p^{(h)}+M_{a}^{(h)}N_b^{(h)}N_c^{(h)}q^{(h)}
   +\frac{i^{(h)}}{d_b^{(h)}}N_c^{(h)}+\delta_{ca}^{(h)}+\delta_{ab}\,,
\end{align}
where $p^{(h)}$ and $q^{(h)}$ are components of three dimensional lattice vectors to be specified shortly, and  $M_\alpha^{(h)}=n^h_\alpha-m^h_\alpha$. After integrating $\phi^{j,I_{ca}}\phi^{i,I_{ab}}{\phi^\star}^{k,I_{cb}}$ over $y_1$, $y_3$ and $y_5$ and evaluating the condition (\ref{eq:soldio}), one gets
\begin{align}
   \int_0^1 \int_0^1  \int_0^1 \mathrm{d}y_1\mathrm{d}y_3\mathrm{d}y_5\, \phi^{j,I_{ca}}_0\phi^{i,I_{ab}}_{\delta_{ca}}{\phi^\star}^{k,I_{cb}}_{\delta_{\delta_{cb}}}=
   &\,\,\,  \mathcal{N}^{ca}_j\mathcal{N}^{ab}_i\mathcal{N}^{cb}_k\nonumber\\
   &\hspace*{-15em}\sum_{p\in\Lambda_p^3}
   \sum_{q\in\Lambda_q^3}\prod_{h=1}^3
 \, \mathrm{e}^{\pi i \left(\frac{i^{(h)}}{d_b^{(h)}I_{ab}^{(h)}}
 +\frac{j^{(h)}}{d_a^{(h)}I_{ca}^{(h)}}
 +\frac{k^{(h)}}{d_c^{(h)}I_{cb}^{(h)}}+2q^{(h)}\right)^2K_h\left| I_{ab}^{(h)}I_{bc}^{(h)}I_{ca}^{(h)}\right|}
 \label{eq:overlaphalfintegrated}\\
 \nonumber
 &\hspace{-15em}
 \mathrm{e}^{-2\pi\frac{\tilde I_{cb}^{(h)}}{\mathrm{Im}(K_h)}\left[\mathrm{Im}(z_h)+N_{abc}^{(h)}\mathrm{Im}(K_h)p^{(h)}+\left(M_a^{(h)}N_b^{(h)}N_c^{(h)}q^{(h)}+\frac{i^{(h)}}{d_b^{(h)}}\frac{I_{cb}^{(h)}}{\tilde I_{cb}^{(h)}}-\frac{k^{(h)}}{d_c^{(h)}}\frac{I_{ca}^{(h)}}{\tilde I_{cb}^{(h)}}\right)\mathrm{Im}(K_h)\right]^2}
 \,,
\end{align}
where $\Lambda^3_p$ and $\Lambda^3_q$ are three dimensional lattices, 
with a lattice structure 
such that $\lambda_{ab}$, $\lambda_{ca}$ and $\lambda_{cb}$ in (\ref{eq:soldio})  belong to 
the lattices $\Gamma_a\cap\Gamma_b$, $\Gamma_c\cap\Gamma_a$ and $\Gamma_c\cap\Gamma_b$, 
respectively.
The components $p^{(h)}$ have to be chosen, such that the vectors $(N_{abc}^{(1)}p^{(1)},N_{abc}^{(2)}p^{(2)},N_{abc}^{(3)}p^{(3)})^T$ belong to 
the $SO(6)$ lattice. Therefore $\Lambda_p^3$ takes the form
\begin{align}
    &\Lambda_p^3=\text{span}\left(\begin{pmatrix}1\\0\\0 \end{pmatrix},\, \begin{pmatrix}0\\1\\0 \end{pmatrix},\,\begin{pmatrix}0\\0\\1 \end{pmatrix}\right) \quad\text{if $\Gamma_a\cap\Gamma_b\cap\Gamma_c$ is of the form $\Gamma_1$}\,,\nonumber\\
      &\Lambda_p^3=\text{span}\left(\begin{pmatrix}\frac{1}{2}\\1\\0 \end{pmatrix},\, \begin{pmatrix}\frac{1}{2}\\-1\\0 \end{pmatrix},\,\begin{pmatrix}0\\0\\1 \end{pmatrix}\right) \quad\text{if $\Gamma_a\cap\Gamma_b\cap\Gamma_c$ is of the form $\Gamma_2$}\,,\\\nonumber
      &  \Lambda_p^3=\text{span}\left(\begin{pmatrix}\frac{1}{2}\\1\\0 \end{pmatrix},\, \begin{pmatrix}\frac{1}{2}\\-1\\0 \end{pmatrix},\,\begin{pmatrix}0\\1\\1 \end{pmatrix}\right) \quad\text{if $\Gamma_a\cap\Gamma_b\cap\Gamma_c$ is of the form $\Gamma_3$}\,
\end{align}
and the terms $N_{abc}^{(h)}\mathrm{Im}(K_h)p^{(h)}$ in (\ref{eq:overlaphalfintegrated}) are components of vectors in a lattice with fundamental cell $\tilde C$. That means a shift of $p$ 
by one of the generators of $\Lambda_p^3$ in (\ref{eq:overlaphalfintegrated}) can be absorbed into the integration by shifting the domain of integration to a neighbouring parallelepiped in the 
lattice with fundamental cell (\ref{eq:tildeCdomain}). That way, the sum over $p$ can be absorbed 
into the integration by enlarging the domain of integration,
\begin{equation*}
    \int_{\tilde C}\mathrm{d}y_2\mathrm{d}y_4\mathrm{d}y_6
    \sum_{p\in\Lambda_p^3}\ldots \rightarrow  \int_{\mathbb{R}^3}\mathrm{d}y_2\mathrm{d}y_4\mathrm{d}y_6 \ldots \,.
\end{equation*}
Now the remaining integration can be performed,
\begin{align}
&\hspace*{-1.5em} \mathcal{N}_j^{ca} \mathcal{N}_k^{ab}\mathcal{N}_k^{cb}\prod_{k=1}^3
  \frac{\mathrm{Im}{\tau_k}}{\mathrm{Im}(K_k)|1-\tau_k|^2}
  \int_\mathbb{R}\mathrm{d}(\mathrm{Im}(z_k))
\nonumber \\
& \hspace*{1.5em}
\sum_{q\in\Lambda_q^3}\prod_{h=1}^3\mathrm{e}^{\pi \text{i} 
  \left(\frac{i^{(h)}}{d_b^{(h)}I_{ab}^{(h)}}+\frac{j^{(h)}}{d_a^{(h)}I_{ca}^{(h)}}
  +\frac{k^{(h)}}{d_c^{(h)}I_{cb}^{(h)}}+2q^{(h)}\right)^2
  K_h\left||I_{ab}^{(h)}I_{bc}^{(h)}I_{ca}^{(h)}\right|}
 \label{eq:final} \\\nonumber
 &\hspace*{3em}\mathrm{e}^{-2\pi\frac{\tilde I_{cb}^{(h)}}{\mathrm{Im}(K_h)}\left[\mathrm{Im}(z_h)+\left(M_a^{(h)}N_b^{(h)}N_c^{(h)}q^{(h)}+\frac{i^{(h)}}{d_b^{(h)}}\frac{I_{cb}^{(h)}}{\tilde I_{cb}^{(h)}}-\frac{k^{(h)}}{d_c^{(h)}}\frac{I_{ca}^{(h)}}{\tilde I_{cb}^{(h)}}\right)\mathrm{Im}(K_h)\right]^2}=\\
 \nonumber
&\hspace*{+6em} 
\mathcal{N}_j^{ca}\mathcal{N}_k^{ab}\mathcal{N}_k^{cb}\left[\prod_{l=1}^3
\frac{\mathrm{Im}(\tau_l)}{|1-\tau_l|^2}\left(2\tilde I_{cb}^{(l)}\mathrm{Im}(K_l)\right)^{-\frac{1}{2}}\right]\sum_{q\in\Lambda_q^3}
\mathrm{exp}{\left\{-\frac{A_{i,j,k}(q)}{2\pi\alpha^\prime}\right\}}\,,
\end{align}
with
\begin{equation*}
    A_{i,j,k}(q)=-2\pi^2 \alpha^\prime \text{i} \sum_{h=1}^3\left(\frac{i^{(h)}}{d_b^{(h)}I_{ab}^{(h)}}
    +\frac{j^{(h)}}{d_a^{(h)}I_{ca}^{(h)}}
    +\frac{k^{(h)}}{d_c^{(h)}I_{cb}^{(h)}}+2q^{(h)}\right)^2K_h
    \left|I_{ab}^{(h)}I_{bc}^{(h)}I_{ca}^{(h)}\right|\,.
\end{equation*}
matching the definition of 
\cite{Forste:2014bfa}. In \cite{Forste:2014bfa} dimensionful K{\"a}hler moduli
have been used. The explicit relation is (see (\ref{eq:scalefix})) 
\begin{equation*}
4\pi^2 \alpha^\prime\text{Im} K_h = g^{(h)} 
\end{equation*}
where $g^{(h)}$ is the determinant of the metric in the $h^{\text{th}}$ complex plane 
in the coordinates (\ref{eq:canbas}), as it was used in \cite{Forste:2014bfa}.
Up to now the lattice, to which the summation index $q$ belongs, has not 
been specified. It can be deduced from (\ref{eq:soldio}): 
The $p$ dependence in (\ref{eq:soldio}) is eliminated in linear combinations of the three 
Diophantine equations
\begin{align}
\lambda_{ab}^{(h)}-\lambda_{ca}^{(h)}-\delta_{ca}^{(h)}=&
2I_{cb}^{(h)}N_a^{(h)}q^{(h)}+\frac{j^{(h)}}{d_a^{(h)}}N_b^{(h)}
+\frac{k^{(h)}}{d_c^{(h)}}N_a^{(h)}\nonumber\\
\lambda_{ca}^{(h)}-\lambda_{cb}^{(h)}-\delta_{ca}^{(h)}-\delta_{ab}^{(h)}
=&2I_{ba}^{(h)}N_c^{(h)}q^{(h)}-\frac{i^{(h)}}{d_b^{(h)}}N_c^{(h)}
-\frac{k^{(h)}}{d_c^{(h)}}N_a^{(h)}\label{eq:soldio2}\\\nonumber
\lambda_{cb}^{(h)}-\lambda_{ab}^{(h)}-\delta_{ab}^{(h)}
=&2I_{ac}^{(h)}N_b^{(h)}q^{(h)}-\frac{j^{(h)}}{d_a^{(h)}}N_b^{(h)}
-\frac{i^{(h)}}{d_b^{(h)}}N_c^{(h)}\,.
\end{align}
The left-hand sides of (\ref{eq:soldio2}) are components of $\Lambda_{\text{SO(6)}}$ 
lattice vectors. Renaming the summation index $2q^{(h)}\rightarrow\ell^{(h)}$, 
this leads to the 
following conditions ($h\in\left\{ 1,2,3\right\}$)
\begin{align}
  I_{cb}^{(h)}N_a^{(h)}\ell^{(h)}+\frac{j^{(h)}}{d_a^{(h)}}N_b^{(h)}
  +\frac{k^{(h)}}{d_c^{(h)}}N_a^{(h)}&\in\mathbb{Z}\,,\nonumber\\
  I_{ba}^{(h)}N_c^{(h)}\ell^{(h)}-\frac{i^{(h)}}{d_b^{(h)}}N_c^{(h)}
  -\frac{k^{(h)}}{d_c^{(h)}}N_a^{(h)}&\in\mathbb{Z}\,,
  \label{selection1}\\
  I_{ac}^{(h)}N_b^{(h)}\ell^{(h)}-\frac{j^{(h)}}{d_a^{(h)}}N_b^{(h)}
  -\frac{i^{(h)}}{d_b^{(h)}}N_c^{(h)}&\in\mathbb{Z}\,,
\nonumber
\end{align}
and
\begin{align}
 \sum_{h=1}^3 I_{cb}^{(h)}N_a^{(h)}\ell^{(h)}
 +\frac{j^{(h)}}{d_a^{(h)}}N_b^{(h)}+\frac{k^{(h)}}{d_c^{(h)}}N_a^{(h)}&=0\mod 2\,,
 \nonumber\\
 \sum_{h=1}^3 I_{ba}^{(h)}N_c^{(h)}\ell^{(h)}-\frac{i^{(h)}}{d_b^{(h)}}N_c^{(h)}
 -\frac{k^{(h)}}{d_c^{(h)}}N_a^{(h)}&=0\mod 2\,,
 \label{selection2}\\
  \sum_{h=1}^3I_{ac}^{(h)}N_b^{(h)}\ell^{(h)}-\frac{j^{(h)}}{d_a^{(h)}}N_b^{(h)}
  -\frac{i^{(h)}}{d_b^{(h)}}N_c^{(h)}&=0\mod 2\,.
\nonumber
\end{align}
Inserting the normalisation factors (\ref{eq:normcond}) into (\ref{eq:final}), 
the Yukawa couplings take the form
\begin{equation}
   |\lambda_{ijk}|= {\alpha^\prime}^{3/2}\text{e}^{\Phi_b/2}
   \prod_{k=1}^3\sqrt{\frac{|1-\tau_k|^2}{\text{Im}\tau_k}}\left(2\mathrm{Im}(K_h)
   \frac{\tilde I_{ab}^{(h)}\tilde I_{ca}^{(h)}}{\tilde I_{cb}^{(h)}}\right)^\frac{1}{4}\sum_{\ell\in\Lambda^3_\ell}\mathrm{exp}
   \left\{-\frac{A_{i,j,k}(\ell)}{2\pi\alpha^\prime}\right\}\,,
\label{eq:yuk}
\end{equation}
with
\begin{equation*}
    A_{i,j,k}(\ell)=-2\pi^2\alpha^\prime \text{i}
    \sum_{h=1}^3\left(\frac{i^{(h)}}{d_b^{(h)}I_{ab}^{(h)}}
    +\frac{j^{(h)}}{d_a^{(h)}I_{ca}^{(h)}}
    +\frac{k^{(h)}}{d_c^{(h)}I_{cb}^{(h)}}
    +\ell^{(h)}\right)^2K_h|I_{ab}^{(h)}I_{bc}^{(h)}I_{ca}^{(h)}|\,
\end{equation*}
$\ell\in\Lambda^3_\ell$ satisfying the selection rules in (\ref{selection1}) and (\ref{selection2}).
In the T-dual type IIA  setting, the Yukawa couplings were computed in \cite{Forste:2014bfa},
\begin{equation*}
    \left|\lambda^{(IIA)}_{ijk}\right| =h_{\text{qu}}\sum_{\ell\in\Lambda^3_\ell}\mathrm{exp}
   \left\{-\frac{A_{i,j,k}(\ell)}{2\pi\alpha^\prime}\right\}\,,
\end{equation*}
where the computation of $h_{\text{qu}}$ has not been performed. A direct calculation should be possible e.g.\ along the lines of \cite{Cvetic:2003ch}. Here, as in \cite{Cremades:2004wa}, 
its leading behaviour, in the small angle limit, will be deduced by T-dualising 
back the type IIB classical calculation. For easier comparison to the factorisable case
the following abbreviations are useful
\begin{align}
    {\mathcal A}^{(h)} & = 4\pi^2\alpha^\prime\frac{\text{Im}\tau_h}{\left|1-\tau_h\right|^2}\,\,\,\text{such\ that}\,\,\, \text{Volume}\left( T^6_{\text{IIB}}\right) = 
    \prod_{h=1}^3{\mathcal A}^{(h)}  ,\\
    \theta_{ab}^{(h)} & = 4\pi \frac{\tilde{I}_{ab}}{{\mathcal A}^{(h)}/\alpha^\prime} .
\end{align}
Taking into account also the dilaton shift (\ref{eq:expdils}) and using (\ref{eq:scalefix}) to obtain a manifestly dimensionless coupling, one finds
\begin{equation*}
    h_{\text{qu}} = \frac{\text{e}^{\Phi_a/2}}{\left( 2\pi\right)^{9/4}} 
    \prod_{h=1}^3 \left(\frac{\theta_{ab}^{(h)}\theta_{ca}^{(h)}}
    {\theta_{cb}^{(h)}}\right)^\frac{1}{4} .
\end{equation*}
This result looks exactly as the one reported in \cite{Cremades:2004wa} for 
factorisable
tori. Here, however the definition of $\theta_{\alpha\beta}^{(h)}$ has been 
modified through a 
modified ${\cal A}^{(h)}$ and $\tilde{I}_{\alpha\beta}^{(h)}$. 
The meaning is the same; in type IIB diluted flux implies small
$\theta^{(h)}_{\alpha\beta}$'s wich in type IIA yield
the level spacing in the quantised  open string stretching from 
brane $\alpha$ to brane $\beta$.

 \section{Conclusions}
  
In the present paper, Yukawa couplings where computed along the lines of \cite{Cremades:2004wa}. However, here a particular non factorisable six-torus was
considered. This arose as a T-dual of a torus generated by the SO(12) root lattice. For cases in which the SO(12) root lattice is replaced by another sublattice of a factorisable lattice straightforward modifications of the presented calculations are expected. Compared to \cite{Cremades:2004wa}, however, some less straightforward 
adjustments had to be performed. Gauge indices as well as zero mode labels take values in quotient lattices which appear as generalisations of products of finite sets of integers. On the type IIA side an SO(12) lattice playing a role in labelling the intersection points was directly related to the compactification lattice. In the T-dual
description, this SO(12) lattice shows up in a rather indirect way when labelling zero modes. For non coprime flux ranks, not all components of a chiral multiplet are related by boundary conditions and hence expressed by the same set of zero modes. 

T-dualising back to type IIA one can identify leading contributions to a factor which can be determined only by a quantum computation on the type IIA side. The result looks exactly as in the factorisable case \cite{Cremades:2004wa}, with some straightforward modifications in the 
definitions of variables. To confirm the presented result, one could 
in principle perform T-duality along other cycles. This is expected to be more
complicated since the cycles of the presented calculation have been chosen such that 
they lie within complex planes. 

It would be interesting to investigate to what extend the presented type IIB calculation can be generalised to cases being not T-dual to type IIA models of the considered kind. 
Abelian Wilson lines have not been turned on for simplicity. In the T-dual IIA setting they correspond to an offset from a brane passing through the origin. Their inclusion is expected to be straightforward.  
Finally, of course, applications to actual model building would  be nice. The presented
configuration generalises known cases and might help accommodating desirable phenomenological aspects.
\section*{Acknowledgements}
We thank Josua Faller for collaboration at an early stage of the presented project. 
This work was supported by SFB-Transregio TR33 ``The Dark Universe'' (Deutsche
Forschungsgemeinschaft) and ``Bonn Cologne Graduate School for Physics and 
Astronomy'' (BCGS). 
\begin{appendix}
\section{Quotient lattices, divisors and multiples of 
integral matrices \label{ap:proof}}  
As discussed in (\ref{eq:latmat}) lattices will be associated to
integral three by three matrices: $\Gamma_a$ to $A$, $\Gamma_b$ to
$B$, $\Gamma_d$ to $D$, and $\Gamma_a \cap \Gamma_b$ to $M$. With the
following definitions one can establish relations among these
matrices. 
\vspace*{0.1in}

\noindent{\bf Definition:} Let $A$, $B$, $D$ be integral
matrices. Then $D$ is a {\it left 
  divisor} of $A$ if there is an integral matrix $M_a$ such that $A= D
M_a$. Further, $D$ is the {\it greatest common left divisor} of $A$ and $B$
if it is a left divisor of $A$ and $B$ and any other left divisor of
$A$ and $B$ is a left divisor of $D$. 

\vspace*{0.1in}

Clearly, the matrix $D$ containing generators of $\Gamma_d$ is a
greatest common left divisor of $A$ and $B$. 
The greatest common left divisor is unique up to multiplication by
unimodular matrices which corresponds to choosing an equivalent set of
lattice generators (see e.g.\ \cite{Newman1972}). An explicit
construction in terms of matrices taking the three by six matrix
$\left(A,B\right)$ to its Smith normal form can be found in
\cite{Bachem1976} (proof of Proposition 3.4) where  
the existence of the matrices $P$ and $Q$ introduced in (\ref{eq:DAB})
is proven.

Similarly one can identify the matrix $M$ with the lowest common right
multiple of $A$ and $B$. Its definition is:

\vspace*{0.1in}

\noindent{\bf Definition:} The integral matrix $M$ is a {\it right
  multiple} 
of the integral matrix $A$ if there is an integral matrix $N_a$ such
that $M= A N_a$. $M$ is the {\it lowest common right multiple} of the
integral matrices $A$ and $B$ if it is a right multiple of $A$ and 
$B$ and any other right multiple of $A$ and $B$ is a right multiple of
$M$. 

\vspace*{0.1in}

The lowest common right multiple $M$and the greatest common left
divisor $D$ have been related in theorem 5 of \cite{thompson1985},
\begin{equation}
M = A D^{-1} B .
\label{eq:lcrm}
\end{equation}
The index of a quotient lattice is related to the integral matrices of
generators as follows. Let $\Lambda_c$ be a sublattice of $\Lambda_l$. Further
$C$ and $L$ denote integral matrices of the corresponding
generators. Then the index of the quotient lattice is
\begin{equation*}
\left| \frac{\Lambda_l}{\Lambda_c}\right| = \left|\frac{\det C}{\det L}\right| .
\end{equation*}
Hence, taking the determinat of (\ref{eq:lcrm}) proves the second
equality in (\ref{eq:indexrange}).  

In the following, examples, relevant for the present paper, will be
listed. As discussed in section \ref{sec:labelling} there are three
possible lattices for $\Gamma_a$ or $\Gamma_b$
\begin{align*}
\Gamma_1 & = \bigotimes_{l=1}^3 N_x^{(l)}{\mathbb Z}\,\,\, ,\,\,\,
           \text{all $N_x^{(l)}$ even} , \\
\Gamma_2 & = \text{span}\left(\left( \begin{array}{c}\frac{N_x^{(1)}}{2}\\
N_x^{(2)}\\ 0\end{array}\right) , \left( \begin{array}{c}\frac{N_x^{(1)}}{2}\\
-N_x^{(2)}\\ 0\end{array}\right) , \left( \begin{array}{c} 0\\0\\
N_x^{(3)}\end{array}\right)\right) \,\,\, \text{$\frac{N_x ^{(1)}}{2}$, $N_x^{(2)}$ odd
and $N_x^{(3)}$ even} ,\\
\Gamma_3 & =   \text{span}\left(\left( \begin{array}{c}\frac{N_x^{(1)}}{2}\\
N_x^{(2)}\\ 0\end{array}\right) , \left( \begin{array}{c}\frac{N_x^{(1)}}{2}\\
-N_x^{(2)}\\ 0\end{array}\right) , \left( \begin{array}{c} 0\\N_x^{(2)}\\
N_x^{(3)}\end{array}\right)\right) \,\,\, \text{$\frac{N_x ^{(1)}}{2}$, $N_x^{(2)}$ 
and $N_x^{(3)}$ odd,}
\end{align*} 
where $x$ stands for $a$ or $b$, respectively. There are six
inequivalent configurations corresponding to symmetric pairings of
these lattices. 

\vspace*{1ex}

\noindent {\bf $\mathbf{\Gamma_a = \Gamma_b=\Gamma_1}$:}
\begin{align*}
&\Gamma_d = \bigotimes_{l=1}^3 d^{(l)} {\mathbb Z},\,\,\, \text{with}\,\,\,
  d^{(l)} = g.c.d.\left( N_a^{(l)}, N_b^{(l)}\right) , \,\,\, 
\left|\frac{\Lambda_{\text{SO(6)}}}{\Gamma_d}\right| =
 \frac{d^{(1)}d^{(2)}d^{(3)}}{2} ,
  \\
&
\Gamma_a
  \cap \Gamma_b = \bigotimes_{l=1}^3 
  \frac{N_a^{(l)} N_b^{(l)}}{d^{(l)}} {\mathbb Z} \,\,\, ,\,\,\,
\left|
 \frac{\Lambda_{\text{SO(6)}}}{\Gamma_a \cap \Gamma_b}\right|
 =\frac{1}{2}\prod_{l=1}^3\frac{N_a^{(l)}N_b^{(l)}}{d^{(l)}} , \\
& \left| \frac{\Lambda_{\text{SO(6)}}}{\Gamma_a}\right| =
\frac{N_a^{(1)}N_a^{(2)}N_a^{(3)}}{2} \,\,\, ,\,\,\, 
\left| \frac{\Lambda_{\text{SO(6)}}}{\Gamma_b}\right|  =  
\frac{N_b^{(1)}N_b^{(2)}N_b^{(3)}}{2} .
\end{align*}

\vspace{1ex}

\noindent{\bf $\mathbf{\Gamma_a = \Gamma_1}$, $\mathbf{\Gamma_b =
    \Gamma_2}$:}
\begin{align*}
&\Gamma_d = \text{span}\left(
  \left(\begin{array}{c}\frac{d^{(1)}}{2}\\
d^{(2)}\\0\end{array}\right),\left( \begin{array}{c}\frac{
                                      d^{(1)}}{2}\\ -d^{(2)}
                                      \\0\end{array}\right)
  ,\left( \begin{array}{c} 0\\0\\d^{(3)}\end{array}\right) \right)
  \,\,\, ,\,\,\, 
\left|\frac{\Lambda_{\text{SO(6)}}}{\Gamma_d}\right| =
 \frac{d^{(1)}d^{(2)}d^{(3)}}{2} , \\ &
\Gamma_a
  \cap \Gamma_b = \bigotimes_{l=1}^3 
  \frac{N_a^{(l)}N_b^{(l)}}{d^{(l)}} {\mathbb Z} \,\,\,,\,\,\, 
\left|
 \frac{\Lambda_{\text{SO(6)}}}{\Gamma_a \cap \Gamma_b}\right|
 =\frac{1}{2}\prod_{l=1}^3\frac{N_a^{(l)}N_b^{(l)}}{d^{(l)}},
\\
& \left| \frac{\Lambda_{\text{SO(6)}}}{\Gamma_a}\right| =
\frac{N_a^{(1)}N_a^{(2)}N_a^{(3)}}{2} \,\,\, ,\,\,\, 
\left| \frac{\Lambda_{\text{SO(6)}}}{\Gamma_b}\right|  =  
\frac{N_b^{(1)}N_b^{(2)}N_b^{(3)}}{2} .
\end{align*}

\vspace*{1ex}

\noindent {\bf $\mathbf{\Gamma_a = \Gamma_1}$, $\mathbf{\Gamma_b = \Gamma_3}$:}
\begin{align*}
&\Gamma_d = \text{span}\left(
  \left(\begin{array}{c}\frac{d^{(1)}}{2}\\
d^{(2)}\\0\end{array}\right),\left( \begin{array}{c}\frac{
                                      d^{(1)}}{2}\\ -d^{(2)}
                                      \\0\end{array}\right)
  ,\left( \begin{array}{c} 0\\d^{(2)}\\d^{(3)}\end{array}\right) \right)
  \,\,\, ,\,\,\, 
\left|\frac{\Lambda_{\text{SO(6)}}}{\Gamma_d}\right| =
 \frac{d^{(1)}d^{(2)}d^{(3)}}{2} , \\ &
\Gamma_a
  \cap \Gamma_b =
 \bigotimes_{l=1}^3 
  \frac{N_a^{(l)}N_b^{(l)}}{d^{(l)}} {\mathbb Z} \,\,\, ,\,\,\,
\left|
 \frac{\Lambda_{\text{SO(6)}}}{\Gamma_a \cap \Gamma_b}\right|
 =\frac{1}{2}\prod_{l=1}^3\frac{N_a^{(l)}N_b^{(l)}}{d^{(l)}} ,
\\
& \left| \frac{\Lambda_{\text{SO(6)}}}{\Gamma_a}\right| =
\frac{N_a^{(1)}N_a^{(2)}N_a^{(3)}}{2} \,\,\, ,\,\,\, 
\left| \frac{\Lambda_{\text{SO(6)}}}{\Gamma_b}\right|  =  
\frac{N_b^{(1)}N_b^{(2)}N_b^{(3)}}{2} .
\end{align*}

\vspace*{1ex}

\noindent {\bf $\mathbf{\Gamma_a = \Gamma_b = \Gamma_2}$:}
\begin{align*}
&\Gamma_d = \text{span}\left(
  \left(\begin{array}{c}\frac{d^{(1)}}{2}\\
d^{(2)}\\0\end{array}\right),\left( \begin{array}{c}\frac{
                                      d^{(1)}}{2}\\ -d^{(2)}
                                      \\0\end{array}\right)
  ,\left( \begin{array}{c} 0\\0\\d^{(3)}\end{array}\right)
  \right) 
\,\,\, ,\,\,\, 
\left|\frac{\Lambda_{\text{SO(6)}}}{\Gamma_d}\right| =
 \frac{d^{(1)}d^{(2)}d^{(3)}}{2} ,
\\ & \Gamma_a
  \cap \Gamma_b = \text{span}\!\left( \!\left( \begin{array}{c}
\frac{N_a^{(1)}N_b^{(1)}}{2d^{(1)}}\\\frac{N_a^{(2)}N_b^{(2)}}{d^{(2)}}\\0\end{array}
\right) \! ,\!
\left( \begin{array}{c}
\frac{N_a^{(1)}N_b^{(1)}}{2d^{(1)}}\\-\frac{N_a^{(2)}N_b^{(2)}}{d^{(2)}}\\0\end{array}
\right) \! ,\!
\left(\begin{array}{c}
        0\\0\\\frac{N_a^{(3)}N_b^{(3)}}{d^{(3)}}\end{array}\right)
\!  \right) 
,
\left|\frac{\Lambda_{\text{SO(6)}}}{\Gamma_a \!\cap\! \Gamma_b}\right|
 =\frac{1}{2}\prod_{l=1}^3\frac{N_a^{(l)}N_b^{(l)}}{d^{(l)}} ,\\ &
\left| \frac{\Lambda_{\text{SO(6)}}}{\Gamma_a}\right| =
\frac{N_a^{(1)}N_a^{(2)}N_a^{(3)}}{2}\,\,\, ,\,\,\,  
\left| \frac{\Lambda_{\text{SO(6)}}}{\Gamma_b}\right|  =  
\frac{N_b^{(1)}N_b^{(2)}N_b^{(3)}}{2} .
\end{align*}

\vspace*{1ex}

\noindent {\bf $\mathbf{\Gamma_a = \Gamma_2}$, $\mathbf{\Gamma_b =\Gamma_3}$:}
\begin{align*}
&\Gamma_d = \text{span}\left(
  \left(\begin{array}{c}\frac{d^{(1)}}{2}\\
d^{(2)}\\0\end{array}\right),\left( \begin{array}{c}\frac{
                                      d^{(1)}}{2}\\ -d^{(2)}
                                      \\0\end{array}\right)
  ,\left( \begin{array}{c} 0\\d^{(2)}\\d^{(3)}\end{array}\right)
  \right) 
\,\,\, ,\,\,\, 
\left|\frac{\Lambda_{\text{SO(6)}}}{\Gamma_d}\right| =
 \frac{d^{(1)}d^{(2)}d^{(3)}}{2} ,
\\ & \Gamma_a
  \cap \Gamma_b = \text{span}\!\left( \!\left( \begin{array}{c}
\frac{N_a^{(1)}N_b^{(1)}}{2d^{(1)}}\\\frac{N_a^{(2)}N_b^{(2)}}{d^{(2)}}\\0\end{array}
\right) \! ,\!
\left( \begin{array}{c}
\frac{N_a^{(1)}N_b^{(1)}}{2d^{(1)}}\\-\frac{N_a^{(2)}N_b^{(2)}}{d^{(2)}}\\0\end{array}
\right) \! ,\!
\left(\begin{array}{c}
        0\\0\\\frac{N_a^{(3)}N_b^{(3)}}{d^{(3)}}\end{array}\right)
\!  \right) 
,
\left|\frac{\Lambda_{\text{SO(6)}}}{\Gamma_a \!\cap\! \Gamma_b}\right|
 =\frac{1}{2}\prod_{l=1}^3\frac{N_a^{(l)}N_b^{(l)}}{d^{(l)}} ,\\ &
\left| \frac{\Lambda_{\text{SO(6)}}}{\Gamma_a}\right| =
\frac{N_a^{(1)}N_a^{(2)}N_a^{(3)}}{2}\,\,\, ,\,\,\,  
\left| \frac{\Lambda_{\text{SO(6)}}}{\Gamma_b}\right|  =  
\frac{N_b^{(1)}N_b^{(2)}N_b^{(3)}}{2} .
\end{align*}

\vspace*{1ex}

\noindent {\bf $\mathbf{\Gamma_a = \Gamma_b =\Gamma_3}$:}
\begin{align*}
&\Gamma_d = \text{span}\left(
  \left(\begin{array}{c}\frac{d^{(1)}}{2}\\
d^{(2)}\\0\end{array}\right),\left( \begin{array}{c}\frac{
                                      d^{(1)}}{2}\\ -d^{(2)}
                                      \\0\end{array}\right)
  ,\left( \begin{array}{c} 0\\d^{(2)}\\d^{(3)}\end{array}\right)
  \right) 
\,\,\, ,\,\,\, 
\left|\frac{\Lambda_{\text{SO(6)}}}{\Gamma_d}\right| =
 \frac{d^{(1)}d^{(2)}d^{(3)}}{2} ,
\\ & \Gamma_a
  \cap \Gamma_b = \text{span}\!\left( \!\left( \begin{array}{c}
\frac{N_a^{(1)}N_b^{(1)}}{2d^{(1)}}\\\frac{N_a^{(2)}N_b^{(2)}}{d^{(2)}}\\0\end{array}
\right) \! ,\!
\left( \begin{array}{c}
\frac{N_a^{(1)}N_b^{(1)}}{2d^{(1)}}\\-\frac{N_a^{(2)}N_b^{(2)}}{d^{(2)}}\\0\end{array}
\right) \! ,\!
\left(\begin{array}{c}
        0\\\frac{N_a^{(2)}N_b^{(2)}}{d^{(2)}}
\\\frac{N_a^{(3)}N_b^{(3)}}{d^{(3)}}\end{array}\right)
\!  \right) \!
,
\left|\frac{\Lambda_{\text{SO(6)}}}{\Gamma_a \!\cap\! \Gamma_b}\right|
 =\frac{1}{2}\prod_{l=1}^3\frac{N_a^{(l)}N_b^{(l)}}{d^{(l)}} ,\\ &
\left| \frac{\Lambda_{\text{SO(6)}}}{\Gamma_a}\right| =
\frac{N_a^{(1)}N_a^{(2)}N_a^{(3)}}{2}\,\,\, ,\,\,\,  
\left| \frac{\Lambda_{\text{SO(6)}}}{\Gamma_b}\right|  =  
\frac{N_b^{(1)}N_b^{(2)}N_b^{(3)}}{2} .
\end{align*}
\end{appendix}

\end{document}